\documentclass[longauth]{aa} 
\usepackage{graphicx}
\usepackage{lipsum} 
\usepackage{txfonts}
\usepackage{amssymb}
\usepackage{amsmath}
\begin{document}

   \title{The HIP 79977 debris disk in polarized light\\ }

   \author{N.~Engler\inst{\ref{instch1}} 
   \and H.M.~Schmid\inst{\ref{instch1}} 
   \and Ch.~Thalmann\inst{\ref{instch1}}  
   \and A.~Boccaletti\inst{\ref{instf4}}      
   \and A.~Bazzon\inst{\ref{instch1}} 
   \and A.~Baruffolo\inst{\ref{insti1}}
   \and J.L.~Beuzit\inst{\ref{instf1}}
   \and R.~Claudi\inst{\ref{insti1}} 
   \and A.~Costille\inst{\ref{instf3}}     
   \and S.~Desidera\inst{\ref{insti1}}
   \and K.~Dohlen\inst{\ref{instf3}}     
   \and C.~Dominik\inst{\ref{instnl2}}    
   \and M.~Feldt\inst{\ref{instd1}}      
   \and T.~Fusco\inst{\ref{instf6}}     
   \and C.~Ginski\inst{\ref{instnl3}}  
   \and D.~Gisler\inst{\ref{instd2}}      
   \and J.H.~Girard\inst{\ref{insteso2}}      
   \and R.~Gratton\inst{\ref{insti1}}
   \and T.~Henning\inst{\ref{instd1}}      
   \and N.~Hubin\inst{\ref{insteso1}}   
   \and M.~Janson\inst{\ref{instd1},\ref{insts1}}     
   \and M.~Kasper\inst{\ref{insteso1}}
   \and Q.~Kral\inst{\ref{inste1}}
   \and M.~Langlois\inst{\ref{instf7},\ref{instf3}}      
   \and E.~Lagadec\inst{\ref{instf5}}
   \and F.~M\'{e}nard\inst{\ref{instf1}}   
   \and M.R.~Meyer\inst{\ref{instch1},\ref{instam1}}
   \and J.~Milli\inst{\ref{insteso2}}         
   \and D.~Mouillet\inst{\ref{instf1}}
   \and J.~Olofsson\inst{\ref{instchi1},\ref{instd1},\ref{instchi2}}
   \and A.~Pavlov\inst{\ref{instd1}}
   \and J.~Pragt\inst{\ref{instnl1}}       
   \and P.~Puget\inst{\ref{instf1}} 
   \and S.P.~Quanz\inst{\ref{instch1}}            
   \and R.~Roelfsema\inst{\ref{instnl1}}
   \and B.~Salasnich\inst{\ref{insti1}}
   \and R.~Siebenmorgen\inst{\ref{insteso1}}
   \and E.~Sissa\inst{\ref{insti1}}
   \and M.~Suarez\inst{\ref{insteso1}}
   \and J.~Szulagyi\inst{\ref{instch1}}  
   \and M.~Turatto\inst{\ref{insti1}}   
   \and S.~Udry\inst{\ref{instch2}} 
   \and F.~Wildi\inst{\ref{instch2}}           }

\institute{
ETH Zurich, Institute for Particle Physics and Astrophysics, Wolfgang-Pauli-Strasse 27, 
CH-8093 Zurich, Switzerland, \email{englern@phys.ethz.ch} \label{instch1}
\and
LESIA, Observatoire de Paris, PSL Research University, CNRS, Sorbonne
Universit\'{e}s, UPMC Univ. Paris 06, Univ. Paris Diderot, Sorbonne Paris
Cit\'{e}, 5 place Jules Janssen 92195 Meudon Cedex, France\label{instf4}
\and
INAF – Osservatorio Astronomico di Padova, Vicolo
dell’Osservatorio 5, 35122 Padova, Italy\label{insti1}
\and
Universit\'{e} Grenoble Alpes, CNRS, IPAG, F-38000 Grenoble, France\label{instf1}
\and
Aix Marseille Universit\'{e}, CNRS, LAM (Laboratoire
d’Astrophysique de Marseille) UMR 7326, 13388, Marseille,
France\label{instf3}
\and
Anton Pannekoek Astronomical Institute, University of Amsterdam,
PO Box 94249, 1090 GE Amsterdam, The Netherlands\label{instnl2}
\and
Max-Planck-Institut f\"{u}r Astronomie, K\"{o}nigstuhl 17, 69117
Heidelberg, Germany\label{instd1}
\and
ONERA, The French Aerospace Lab BP72, 29 avenue de la
Division Leclerc, 92322 Ch\^{a}tillon Cedex, France\label{instf6}
\and
Leiden Observatory, Leiden University, P.O. Box 9513, 2300 RA
Leiden, The Netherlands\label{instnl3}
\and
Kiepenheuer-Institut f\"{u}r Sonnenphysik, Schneckstr. 6, D-79104
Freiburg, Germany\label{instd2}
\and
European Southern Observatory, Alonso de Cordova 3107, Casilla
19001 Vitacura, Santiago 19, Chile\label{insteso2}
\and
European Southern Observatory, Karl Schwarzschild St, 2, 85748
Garching, Germany\label{insteso1}
\and
Department of Astronomy, Stockholm University, AlbaNova University Center, 10691 Stockholm, Sweden\label{insts1}
\and
Centre de Recherche Astrophysique de Lyon, CNRS/ENSL
Universit\'{e} Lyon 1, 9 av. Ch. Andr\'{e}, 69561 Saint-Genis-Laval,
France\label{instf7}
\and
Laboratoire Lagrange, UMR7293, Universit\'{e} de Nice Sophia-Antipolis, 
CNRS, Observatoire de la C\^{o}te d'Azur, Boulevard de l'Observatoire, 
06304 Nice, Cedex 4, France\label{instf5}
\and
Department of Astronomy, University of Michigan, 311 West Hall, 1085 S. University Avenue, Ann Arbor, MI 48109, USA\label{instam1}
\and
Instituto de Física y Astronomía, Facultad de Ciencias, Universidad de Valparaíso, Av. Gran Bretaña 1111, Playa Ancha, Valparaíso, Chile\label{instchi1}
\and
NOVA Optical Infrared Instrumentation Group at ASTRON, Oude
Hoogeveensedijk 4, 7991 PD Dwingeloo, The Netherlands\label{instnl1}
\and
Geneva Observatory, University of Geneva, Chemin des Mailettes
51, 1290 Versoix, Switzerland\label{instch2}
\and
ICM nucleus on protoplanetary disks, ``Protoplanetary discs in ALMA Early Science'', Chile
\label{instchi2}
\and
Institute of Astronomy, University of Cambridge, Madingley Road, Cambridge CB3 0HA, UK
\label{inste1}
     }

   \date{Received ...; accepted ...}

\abstract
{Debris disks are observed around 10 to 20~\% of FGK main-sequence stars 
as infrared excess emission. They are important signposts 
for the presence of colliding planetesimals and therefore provide important information about the evolution of 
planetary systems. Direct imaging of such disks reveals 
their geometric structure and constrains their dust-particle properties.}
{We present observations of the known edge-on debris disk around 
HIP 79977 (HD 146897) taken with the ZIMPOL differential polarimeter 
of the SPHERE instrument. We 
measure the observed polarization signal and investigate the 
diagnostic potential of such data with model simulations.}
{SPHERE-ZIMPOL polarimetric data of the 15~Myr-old F star HIP 79977 
(Upper Sco, 123 pc) were taken in the Very Broad 
Band (VBB) filter ($\lambda_c=735$~nm, $\Delta\lambda=290$~nm) with
a spatial resolution of about 25~mas. Imaging
polarimetry efficiently suppresses  the residual speckle noise from
the AO system and provides a differential signal with
relatively small systematic measuring uncertainties. We 
measure the polarization flux along and perpendicular to the disk spine 
of the highly inclined disk for projected separations between 
$0.2''$ (25 AU) and $1.6''$ (200 AU). 
We perform model calculations for the polarized flux 
of an optically thin debris disk which are used to
determine or constrain the disk parameters of HIP 79977.}
{We measure a polarized flux contrast ratio for the disk of $(F_{\rm pol})_{\rm disk}/F_{\rm \ast}= (5.5 \pm 0.9) \cdot 10^{-4}$ in the
VBB filter. The surface brightness of the 
polarized flux reaches a maximum  
of ${\rm SB}_{\rm max}=16.2\,{\rm mag}\,{\rm arcsec}^{-2}$
at a separation of $0.2''-0.5''$ along the disk spine
with a maximum surface brightness contrast of $7.64\,{\rm mag}\,{\rm arcsec}^{-2}$. The polarized 
flux has a minimum near the star $<0.2''$ because no or only 
little polarization is produced by forward or backward scattering 
in the disk section lying in front of or behind the star. The width 
of the disk perpendicular to the spine shows a systematic increase 
in FWHM from $0.1''$ (12 AU) to $0.3'' - 0.5''$, when going from a 
separation of $0.2''$ to $>1''$. This can be explained by 
a radial blow-out of small grains.  
The data are modelled as a circular dust belt with a well defined
disk inclination $i=85(\pm 1.5)^\circ$ and a radius between $r_0=60$ 
and 90~AU. The radial density dependence is  
described by $(r/r_0)^{\alpha}$ with a steep (positive) power law 
index $\alpha=5$ inside $r_0$ and a more shallow (negative) index 
$\alpha=-2.5$ outside $r_0$. The scattering asymmetry factor lies 
between $g=0.2$ and 0.6 (forward scattering) adopting
a scattering-angle dependence for the fractional polarization
such as that for Rayleigh scattering. }  
{Polarimetric imaging with SPHERE-ZIMPOL of the edge-on debris disk
around HIP 79977 provides accurate profiles for the polarized flux. 
Our data are qualitatively very similar to the case of 
AU Mic and they confirm that edge-on debris disks have a polarization minimum
at a position near the star and a maximum near the projected 
separation of the main debris belt. The comparison of the polarized flux contrast ratio 
$(F_{\rm pol})_{\rm disk}/F_{\rm \ast}$ with the fractional infrared excess provides strong constraints 
on the scattering albedo of the dust.}



\authorrunning{Engler et al.}

\titlerunning{HIP79977 debris disk in polarized light
}

   \maketitle
%

\section{Introduction}

Many main-sequence stars with circumstellar 
dust have been identitified based on the detection of 
infrared (IR) excess emission \citep{Aumann1984, Oudmaijer1992}. For nearby systems with
strong IR excess, like $\beta$ Pic, Fomalhaut, 
HR 4796A and others, it was shown with high contrast 
observations that this dust is located in disks or 
rings \citep{Smith1984, Backman1993, Schneider1999,Kalas2005} around the central star. The
dust is attributed to dust debris from collisions of 
solid bodies in a planetesimal disk,
similar to the Kuiper belt in the solar system 
\citep[see e.g.,][for a review]{Wyatt2008}.
The lifetime of small dust particles, which are the
main component for the IR-excess emission, is very
short because they are blown out of the system by
radiation pressure or stellar winds and therefore
they must be replenished by ongoing collisions in the
system. Bright debris disks are particularly frequent
around young stars where they are the last phase
of the evolution of planet-forming disks and for
this reason young, bright giant planets are often found
in systems with debris disks 
\citep[e.g.,][]{Kalas2008, Marois2008,Lagrange2010}. For older
stars ($>10^8$ yr) the debris disks are rare and usually
faint with a few interesting exceptions which could
be caused by a strong transient collisional event. Debris-disk structure has the potential to reveal the dynamics of
planetary systems and provide very important information
 about their evolution. 

Important aspects for an understanding of the parent bodies
responsible for the debris dust are the disk geometry and
the dust particle sizes, structures, and compositions. The
determination of the geometry requires spatially resolved 
observations of the disk. This can be achieved with 
IR-observations of the thermal emission of the dust 
\citep[e.g.,][]{Stapelfeldt2004, Su2005, Wahhaj2007}, 
or with high-contrast observations of 
the scattered stellar light \citep[e.g.,][]{Golimowski2006, Schneider2014}. 
Particle properties are difficult to derive 
observationally, because the measurements are indirect and
often ambiguous. Typical particle
sizes may be inferred from the spectral energy distribution
in the IR and the separation of the dust from the star. 
For hot dust, the composition can sometimes be inferred 
from spectral features, 
mainly the silicate bands around 10~$\mu$m and 18~$\mu$m 
\citep[e.g.,][]{Chen2006, Duchene2014, Mittal2015, Olofsson2009, Moor2009, Olofsson2012} 
and the color of the scattered light might also indicate
grain size, porosity or composition of the particle \citep[e.g.,][]{Debes2008, Debes2013}. 

Up to now, most high-resolution and high-contrast images
of debris disks in scattered light have been taken with the 
Hubble Space Telescope (HST) or adaptive optics (AO) observation
using large telescopes from the ground. HST is a powerful
high-contrast instrument because the point spread 
function (PSF) is not affected by a turbulent atmosphere
and therefore it provides well calibrated intensity images of
extended disks. AO observations from the ground provide a 
high spatial resolution but they suffer from the variable PSF which
depends strongly on atmospheric conditions. To reveal faint debris disks, 
high-contrast data-reduction techniques like angular differential
imaging (ADI) or reference PSF subtraction must be applied.
This can be particularly difficult for ground-based AO data.

In this work we present data of the debris disk HIP 79977
which was observed with differential polarimetric imaging using
the new, extreme AO instrument SPHERE-ZIMPOL at the VLT \citep{Beuzit2008}. 
Polarimetry is an alternative and very sensitive differential 
measuring method for accurate measurements of the polarized and therefore scattered 
light from circumstellar dust in the bright halo of unpolarized light
from the central star. The measured polarization signal
contains additional diagnostic information on the scattering dust, 
different from the intensity signal. But the diagnostic potential
of polarimetry has hardly been investigated for debris disks because
only a few systems have been observed with polarimetry up until a few 
years ago \citep{Gledhill1991, Tamura2006, Graham2007, Hinkley2009}.
With the advent of new extreme AO systems, such as SPHERE and Gemini Planet Imager (GPI), 
with sensitive polarimetric modes \citep[e.g.,][]{Perrin2015, Olofsson2016, Draper2016}
this technique will become much more attractive. Our data on   
HIP 79977 are also used to demonstrate the capabilities of SPHERE-ZIMPOL for debris 
disks with imaging and polarimetric imaging. 
Therefore, we provide more extensive information on data reduction, analysis, and modeling.

HIP 79977 is a young, 15~Myr old \citep{Pecaut2012},
F2/3V star of the Upper Scorpius association, located at 
a distance of 123$^{+18}_{-14}$ pc \citep{vanLeeuwen2007}. The $\sim$1.5 $M_{\odot}$ 
star is not known to have stellar or planetary companions so far. The infrared excess was detected by the \textit{IRAS} satellite and was associated with a bright debris disk based on the 24~$\mu$m and 70~$\mu$m excesses measured with Spitzer Multiband Imaging Photometer (MIPS) \citep{Chen2011}. The authors supported their suggestion with the high-resolution optical spectra obtained with Magellan MIKE spectrograph which showed no signs of active accretion onto the star. There is not much gas in the disk because only 
a tentative detection of the CO gas was reported by \cite{Lieman-Sifry2016}, 
suggesting that the amount of gas in the disk is 
small compared to the amount of dust. The fractional IR luminosity of 
$L_{\mathrm{IR}}/L_{\star}=5.21\cdot10^{-3}$ of this target is 
high but not exceptional. Among 46 young F-type stars of the Scorpius-Centaurus OB Association with mass $\sim$1.5 $M_{\odot}$ and age between 10 and 17 Myr which were identified as debris disk systems, 11 show a fractional IR luminosity higher than $10^{-3}$ \citep{Jang-Condell2015}. \\
The disk around HIP 79977 was imaged in scattered light intensity, or Stokes $I$, in the H-band 
and also detected with polarimetry
with the Subaru HiCiao instrument \citep{Thalmann2013}. The
observations revealed an edge-on disk extending out
to approximately 2$''$ (250 AU), though its inner regions ($r<0.4''$) were hidden by 
residual speckles. 
These data show that HIP 79977 is a good case for an 
edge-on debris disk fitting well onto the detector
field of view ($3.6''\times 3.6''$) of the SPHERE-ZIMPOL instrument.
Similar full disk observations are not possible with this instrument for the famous 
nearby examples $\beta$ Pic or AU Mic, because they are
too extended. 

The paper is organized as follows. 
In Sect.\,\ref{Observations} we describe the observations 
and present the data. Section \ref{data} is dedicated to the 
methods of the data reduction and Sect. \ref{Data Analysis} 
to the polarimetric data analysis. Then, in Sect. \ref{Modeling}, 
we give a description of our model for the spatial distribution 
of the dust developed to reproduce the morphology of the HIP 79977 
debris disk and present the results of the modeling. Finally, 
in Sect. \ref{Discussion}, we compare results from this work with 
the disk models obtained in previous studies of HIP 79977 and 
discuss the diagnostic potential of polarimetric measurements of debris disks.


\begin{table*} 
      \caption[]{Summary of observations.}
         \label{Settings}
  		\renewcommand{\arraystretch}{1.3}
         \begin{tabular}{ccccccccccc}
            \hline 
            \hline
            Date/Observation&Instrument&Filter&Filter&\multicolumn{3}{c}{Integration Time  [s]}&& \multicolumn{3}{c}{Observing Conditions} \\\cline{5-7} \cline{9-11}
            identification& mode &arm 1&arm 2&DIT$^{1}$&Tot$^{2}$& Eff$^{3}$&& Airmass & Seeing [$''$] & $ \tau_0 $ [ms] \\
            \hline
            \hline
            \noalign{\smallskip}
            \multicolumn{1}{l}{2014-08-15/}\\
            OBS227\_0003-0006&imaging &VBB&I\_PRIM & 60 &2400&1740&&1.00--1.01&0.9--1.7&1.7--2.8\\[5pt]
            \multicolumn{1}{l}{2015-04-24/}\\
            OBS114\_0122-0200& SP & VBB & VBB & 16 &5120& 3872 &&1.03--1.27&1.1--2.2&0.9--1.8\\
           
	   \noalign{\smallskip}
            \hline
            \hline
            \noalign{\smallskip}
         \end{tabular}\\

$^1$ Detector integration time (DIT).\\ $^2$ Total integration time on source.\\  $^3$ Total integration time of all frames used in the data reduction.\\ 
   \end{table*}

\section{Observations} \label{Observations}

The SPHERE ``Planet Finder'' instrument for high-contrast
observations in the near-IR and visual spectral range consists of
an extreme adaptive optics (AO) system and three focal
plane instruments for differential imaging \citep{Beuzit2008,
Kasper2012,Dohlen2006,Fusco2014}. The data described in this
work were taken with the ZIMPOL (Zurich Imaging Polarimeter)
subsystem working in the spectral range from 520~nm to 900~nm  
\citep{Schmid2012,Bazzon2012,Roelfsema2010}. The SPHERE-ZIMPOL
configuration provides a spatial resolution of $20-30$~mas
and observing modes for angular differential imaging and
polarimetric differential imaging. The pixel scale of ZIMPOL
is 3.60~mas per pixel and the field of view is $3.6''\times 3.6''$.
ZIMPOL has two camera arms, cam1 and cam2, and data are taken
simultaneously in both arms, each equipped with its own filter 
wheel. 

A special feature of the ZIMPOL detectors is the row masks
covering every second row of the detector which is 
implemented for high-precision imaging polarimetry using 
a polarimetric modulation and on-chip demodulation technique 
\citep{Schmid2012}. A raw frame taken in imaging mode has
only every second row illuminated and the useful data has a format
of $512\times 1024$ pixels where one pixel represents 
$7.2\times 3.6$~mas on the sky. The same format results
from polarimetric imaging for the perpendicular $I_\perp$ 
and parallel $I_\parallel$ polarization signals stored in 
the ``even'' and ``odd'' rows respectively. The advantage of this 
technique is that the images with opposite polarization $I_\perp$ 
and $I_\parallel$ are recorded using the same detector pixels. 
This significantly reduces the differential aberation between 
$I_\perp$ and $I_\parallel$ and flat-fielding issues. 
In the data reduction the $I_\perp$ and $I_\parallel$ frames, each $512\times 1024$ pixels, are extracted. In a later step in the reduction the $512\times 1024$ pixel images are expanded into $1024\times 1024$ pixel images with a flux conserving interpolation so that one pixel in the reduced image corresponds to $3.6\times 3.6$ mas on sky.   

All SPHERE-ZIMPOL observations of HIP 79997 are summarized in Table\,\ref{Settings}.

Imaging observations of HIP 79977 were carried out 
during a SPHERE commissioning run in August 2014 using the VBB or
RI-band filter ($\lambda_c=735$ nm, $\Delta\lambda= 290$ nm) in cam1 and the 
I-band filter ($\lambda_c=790$ nm, $\Delta\lambda= 153$ nm)
in cam2. A sequence of 40 frames with a total exposure time of 40 min
was taken in pupil tracking mode for angular differential imaging 
\citep[ADI;][]{Marois2006}. The atmospheric conditions were strongly 
variable with a seeing between 0.9$''$ and 1.7$''$ and short 
coherence times between 1.7 and 2.8~ms.  

\begin{figure}
   \centering
  \includegraphics[width=8.5cm]{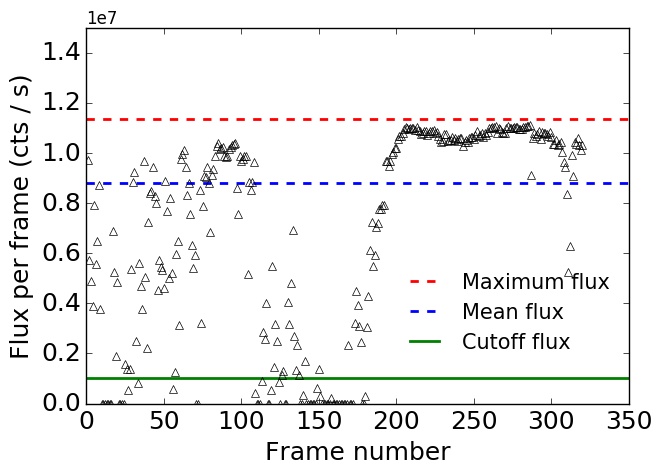}
               \caption{Total counts per second in the frames for the polarimetric 
observations of April 2015 illustrating the impact of clouds on the data.
Essentially only frames with count rates above 
$1\cdot 10^6$ (green line) were used in the data reduction.
The dashed lines mark the maximum counts per frame $1.14\cdot 10^7$ and the 
mean counts $8.6\cdot 10^6$ for the frames considered in the data analysis.
         \label{counts} }
   \end{figure}

Polarimetric measurements were taken as part of the SPHERE
guaranteed time observations (GTO) on April 24, 2015 in field 
stabilized instrument mode (P2) and using the slow polarimetry (SP) detector mode  
with modulation frequency $\sim$27 Hz. 
The wide VBB filters were used in both arms of 
the instrument. We observed the target with four different 
sky orientations on the CCD detectors with position-angle offsets of $0^\circ,\ 50^\circ, 100^\circ$ and $135^\circ$ with respect to sky North. 
We recorded several polarimetric QU-cycles for each position angle. 
In one cycle, the half-wave plate (HWP) is rotated by 0$^\circ$, 
45$^\circ$, 22.5$^\circ$ and 67.5$^\circ$ for measurements of the 
Stokes linear polarization parameters $Q, -Q, U$, and $-U$, respectively. 
In total 320 frames with an on-source integration time of about 85~min were taken. The observing conditions 
for the polarimetric observations were strongly variable with rather poor seeing conditions 
(varying from 1.07$''$ to 2.23$''$) and passing clouds, so that the
AO system loop crashed repeatedly. Figure \ref{counts} shows 
the registered source counts illustrating the variable atmospheric
extinction. 

The peak of the stellar PSF is saturated by 
at most a factor of 10 in the center ($r\leq 3$~pixels) 
for the imaging and also the cloud-free
polarimetric observations. Non-coronagraphic, moderately saturated
observations were chosen to optimize the dynamical range of the data
at small angular separation with not too much sensitivity loss 
at large separation due to read-out noise.

\section{Data reduction}\label{data}
\subsection{Angular differential imaging}
For the basic data reduction steps of 
images of total intensity (Stokes $I$) taken in 2014,
the SPHERE Data Reduction and Handling (DRH) software \citep{Pavlov2008} was used. This includes
the image preprocessing, dark frame subtraction and flat-fielding.
All 40 frames were visually inspected and 11 bad frames 
containing strongly asymmetric PSFs and unexpected
features were rejected (see Table \ref{Settings} for the total effective integration time after frame selection). These effects were caused by phases when
the control loop of the AO system failed or almost failed because
of the ``rough'' atmospheric conditions. To reduce the impact of 
strong PSF variations, all selected frames were rescaled 
by dividing them by the flux measured in an annulus between 
$r_{\rm in}=20$ pixels and $r_{\rm out}=150$ pixels.    

We used a LOCI algorithm
(Locally Optimized Combination of Images, \cite{Lafreniere2007})
to remove the stellar light from the images. LOCI divides each
frame into segmented annuli; for each segment, it
then constructs a matching reference PSF from a linear combination
of similar segments taken from other frames in the dataset.  The
two most important tuning parameters of the algorithm are $N_\delta$
and $N_A$. The former determines the degree to which point sources
in the data are protected from self-subtraction: frames are excluded
from the linear combination if their differential field rotation with
respect to the working frame is so small that a planet located in the
working annulus would move by less than $N_\delta$ times the Full
Width at Half Maximum (FWHM) between the two frames. The second
parameter, $N_A$, describes the size of the region in which the
optimization is performed in units of resolution elements.

When optimized for point-source detection, LOCI causes dramatic
self-subtraction and therefore signal loss in extended structures
such as circumstellar disks.  However, the parameters can be adapted
to preserve more disk flux while still maintaining some of the
algorithm's efficacy at speckle removal (``conservative LOCI'').  Here, we adopt a small
value of $N_\delta=0.5$ and a large value of $N_A= 10000$, which
has proven effective in past studies
\citep[e.g.,][]{Thalmann2010, Buenzli2010, Thalmann2011}.

\begin{figure*}   
    \includegraphics[width=16.5cm, trim=-1cm 0 0 -1.2cm]{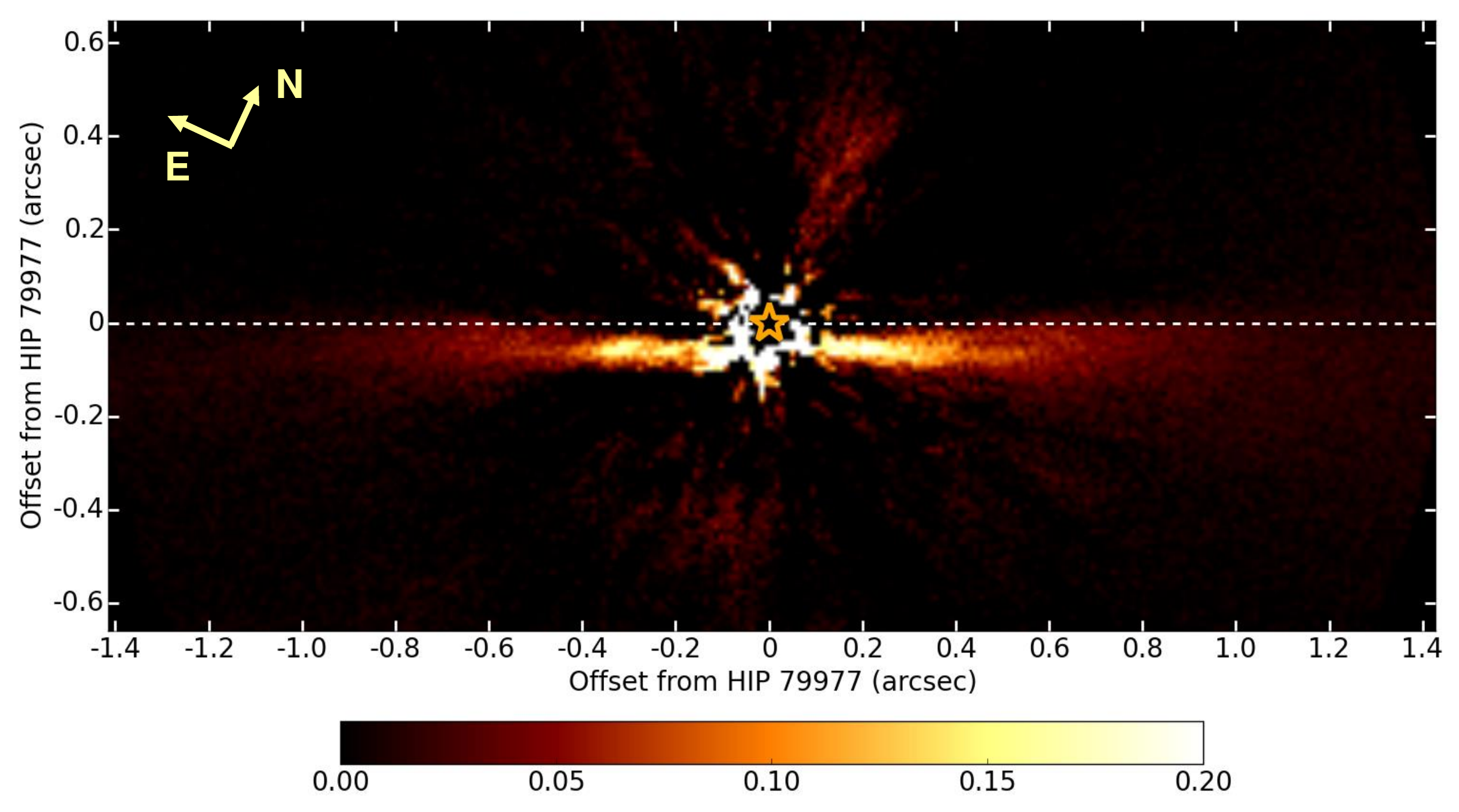} 
               \caption{Composite image of debris disk around HIP 79977 with the VBB and I-band filters obtained with LOCI data reduction. The original data were $3\times 3$ binned to reduce the effect of the noise. The position of the star is marked by an asterisk in orange. The white dotted line shows the position of the expected lines of nodes for an inclined disk ring. 
The color-scale is given in arbitrary units. }
         \label{imaging data}
\end{figure*}

Scattered light from the debris disk is detected in the I-band 
and VBB data along a line oriented in ESE -- WNW direction which
is slightly offset $(<0.1'')$ from the star towards SSW. Emission is
visible from $0.1''$ to beyond $1''$ from the star as shown 
in Fig.~\ref{imaging data}.
The LOCI reduction can be interpreted as an edge-on disk with a high 
inclination $i >80^\circ$. 
At small separations from the star the southwest side of the 
disk is bright while the northeast side is not detected. The main
disk features observed by us confirm the H-band observation of 
\citet{Thalmann2013} but our data provide a higher spatial resolution
and S/N-detection. 

\subsection{Polarimetric differential imaging}  \label{PDI}

The data have been reduced with the SPHERE-ZIMPOL software developed at the ETH Zurich. 
The basic reduction steps are essentially identical to the
SPHERE DRH software. 

The polarimetric data were also visually inspected and correctly recorded frames 
with count rates above $1\cdot 10^6$ were selected for the data reduction. 
The total integration time after removing bad frames is 3872 s (see Table \ref{Settings}).  
 
The ZIMPOL is designed 
as sensitive imaging polarimeter and it includes a series
of differential techniques to reduce systematic effects for
the detection of faint polarimetric signals \citep{Bazzon2012,Thalmann2008}. 
This includes the combination of polarimetric modulation 
and a synchronous on-chip demodulation where opposite
polarization modes $I_\perp$ and $I_\parallel$ are stored with
charge shifting in the ``odd'' and ``even'' detector pixel rows on the CCD. 
Furthermore, every second frame reverses the up and down
shifting to account for charge shifting differences, and every
$Q^+=I_\perp-I_\parallel$-frame is complemented with a 
$Q^-= I_\parallel-I_\perp$-frame to compensate the instrumental 
polarization. These steps are intrinsic parts of the observing
strategy.

A basic data reduction is often sufficient to identify a bright circumstellar 
disk. Sometimes, better results can be obtained if also the residual
telescope polarization is taken into account. This is a more difficult task, because $p_{\rm T}$ and 
the orientation $\theta_{\rm T}$ of this polarization depends on color, 
rotation mode P1 or P2, and pointing direction and the correction 
law is not available yet. A preliminary analysis of the calibration with
zero-polarization standard stars indicates a telescope instrumental polarization 
at the level of $p_{\rm T} \approx 0.5~\%$. A useful work-around provides a forced 
normalization of the total counts of corresponding frames, for example, 
$I_\perp=I_\parallel$ or $Q^+ = Q^-=0$. However, such procedures can 
introduce spurious signals and must be applied with caution because they  
treat the intrinsic polarization of the 
central star or an interstellar polarization signal like a 
(instrumental) telescope polarization signal. 

Early ZIMPOL-SPHERE observations demonstrate that the basic reduction 
steps combined with the forced normalization trick yield high-quality 
polarimetric images of proto-planetary disks \citep{Garufi2016, Stolker2016}. 
However, one should be aware, 
that the contrast of even a bright debris disk like HIP 79977 
is about one order of magnitude lower than a bright proto-planetary 
disk. For this reason additional systematic effects need to be corrected. 

Systematic noise from the instrument can also be reduced by averaging data
taken with different field orientations. We have taken such data 
for HIP 79977 but the improvement is limited because certain 
position angles were strongly affected by clouds. 
Important for the quality of the final result is a careful
centering of individual images to a high precision. This works well with a fit 
of a two-dimensional (2D) Gaussian function to the steep intensity gradients of the 
stellar profile, despite the often saturated central peak. The estimated centering 
accuracy is $<0.3$ pixels or $<1$ mas.

Finally, we found that the combination
of the final frames from cam1 and cam2 is also very beneficial for
the image quality. Spurious polarization signals introduced by 
temporal variations of the atmosphere and AO system are opposite in 
the two channels if the same filters are used in cam1 and cam2 so that
in a mean image some temporal effects are compensated.  

\begin{figure*} 
   \centering
   \includegraphics[width=17cm]{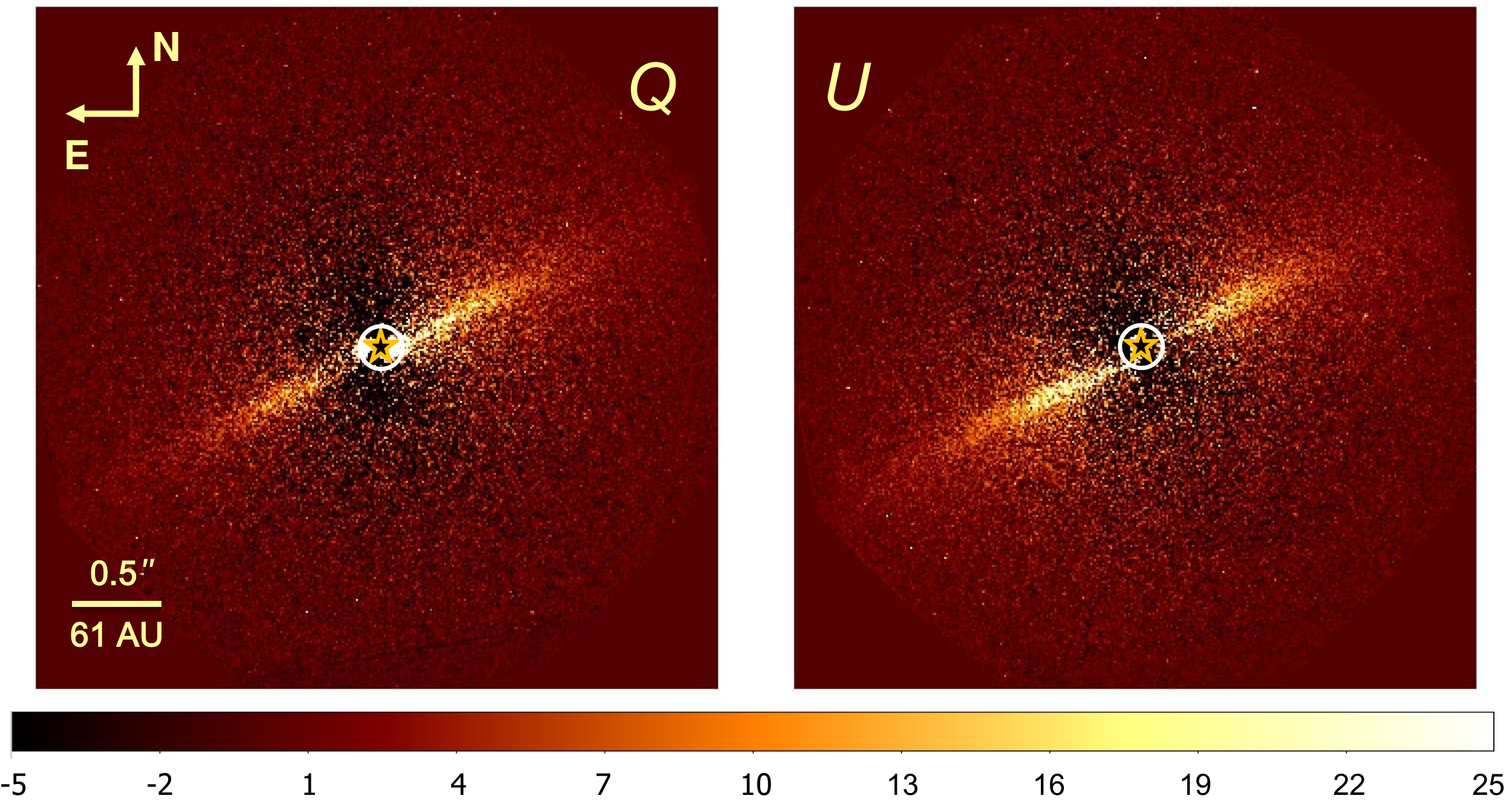}
      \caption{Polarimetric differential imaging data of HIP 79977 with the VBB filter (590-880 nm). The mean images show polarized flux Stokes $Q$ (left) and $U$ (right) after $3\times 3$ binning. The position of the star is marked by an asterisk in orange. The image region located within a white stellarcentric circle with a radius of $\sim 0.12''$ is dominated by the strong speckles variations. The color-bar shows the counts per binned pixel. \label{qu}    }
 \end{figure*}

After all these data reduction steps, significant signals of polarized 
light from the debris disk are clearly visible in the Stokes 
$Q$ and $U$ images (Fig. \ref{qu}). The central star is marked with an asterisk,
and the white circle shows the immediate region surrounding the star which is affected by saturation and strong speckle noise. 

The $Q$ and $U$ images both show a faint negative halo
around the central star. This could be explained by a residual
polarization signal of $-0.3$~\% and $-0.2$~\% of the stellar PSF
in the $Q$ and $U$ images respectively which could be the
result of the applied ``forced normalization'' described above.  
This effect can be corrected by:
\begin{equation}
Q_{new} = Q + 0.003*I_q
\end{equation}
\begin{equation}
U_{new} = U + 0.002*I_u,
\end{equation}
where $I_q$ and $I_u$ are mean stellar intensities measured in $Q$ and $U$ cycles respectively.\\

Diffraction from the telescope spider could be an additional effect
contributing to the observed halo. The orientation of the vertical
telescope spider coincides during the polarimetric observations with
the negative regimes above and below the disk in Q and U images. Further characterization of the instrument is needed
to understand the origin of this signal.

\paragraph{Azimuthal polarization images:} From the Stokes 
\textit{Q} and \textit{U} maps we can compute the intensity 
of the polarized flux $P=\sqrt{Q^2+U^2}$. However, $P$ is affected for 
low signal-to-noise data by a systematic bias effect because of squaring of $Q$ and $U$ parameters.
Therefore we characterize the disk polarization pattern with
a locally defined azimuthal / radial $Q$- and $U$-parameter definition  
with respect to the central light source as discussed in \cite{Schmid2006}. 
Single scattering off dust particles in optically thin debris disks 
generates linearly polarized light with the electric field vector 
azimuthally oriented with respect to the star. 
Polarization in the azimuthal direction is defined by the Stokes 
parameter $ Q_\varphi $:
\begin{equation}
Q_\varphi = - (Q\cos 2\varphi+U\sin 2\varphi),
\end{equation}
where $\varphi$ is the polar angle between north and the point 
of interest measured from the north over east.
The Stokes parameter $U_\varphi$:
\begin{equation}
U_\varphi = -Q\sin 2\varphi+U\cos 2\varphi
\end{equation}
defines the polarization pattern in the directions $\pm45^\circ$ with respect to the $Q_\varphi $ direction.

\begin{figure*}
   \centering
   \includegraphics[width=17cm, trim=-0.5cm 0 0 0cm]{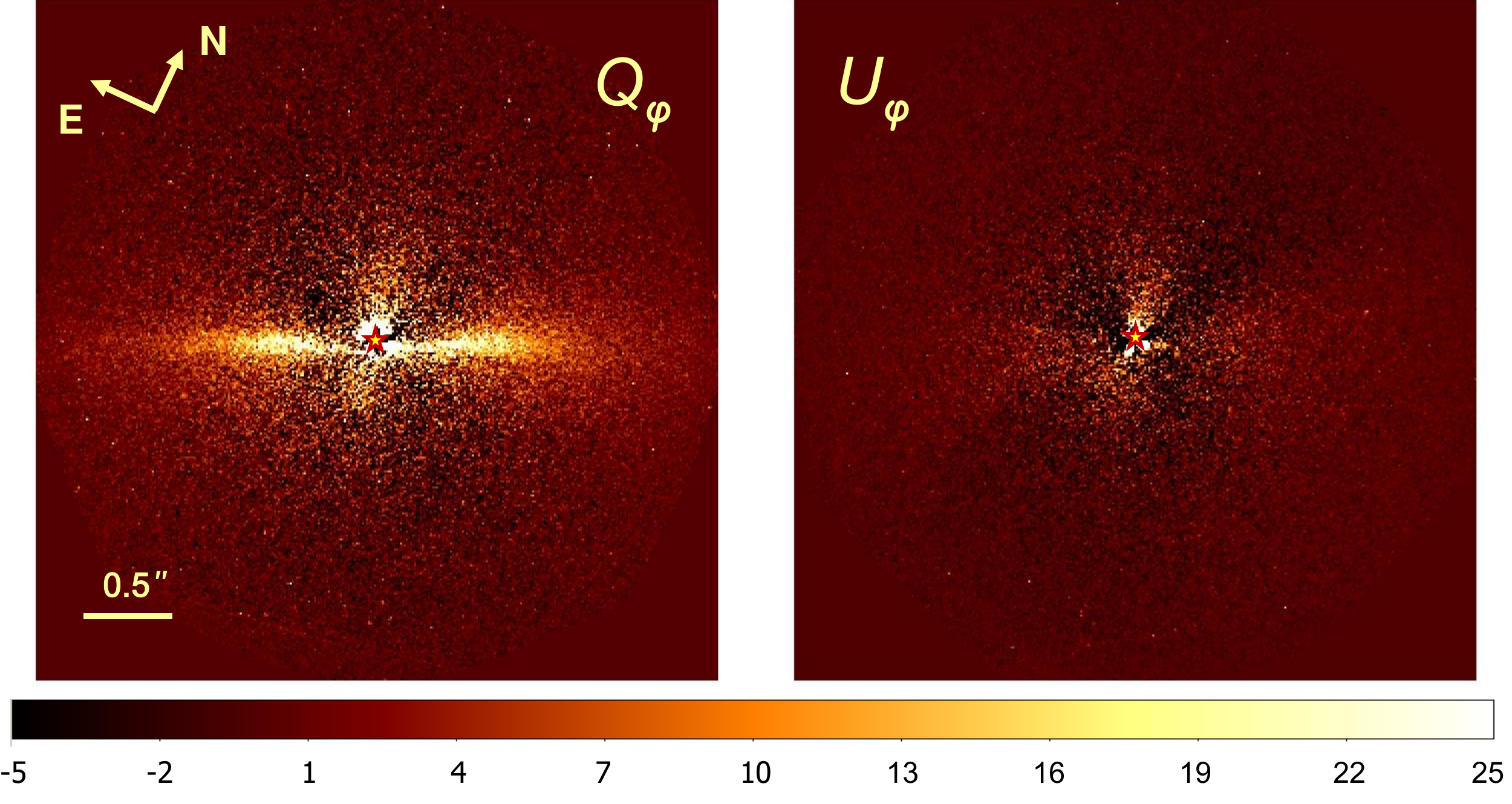}
   \includegraphics[width=17cm, trim=1cm 0 0 -1.2cm]{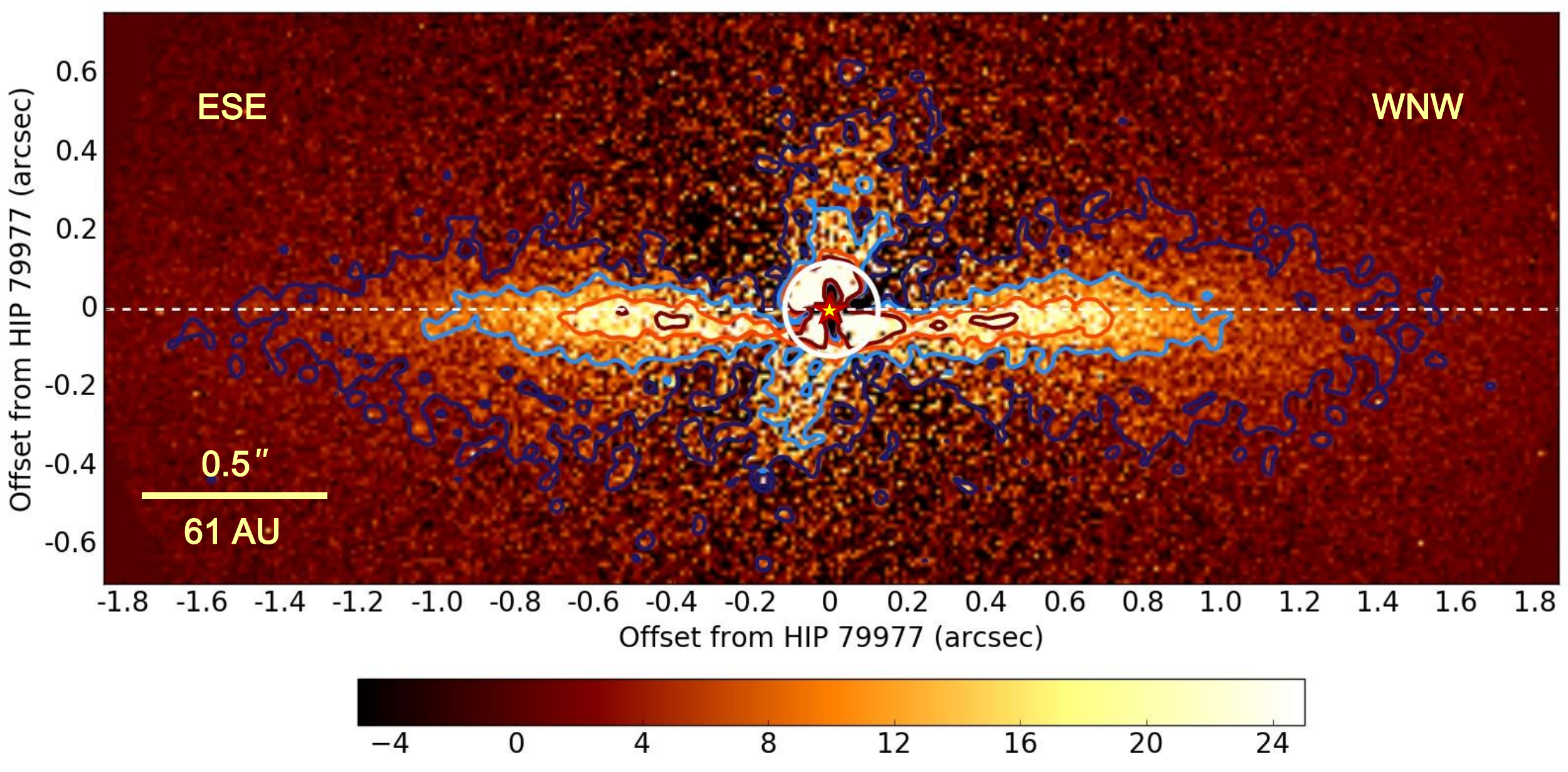} 
         \caption{Polarimetric differential imaging data of HIP 79977 with the VBB filter (590-880 nm). The original data were $3\times 3$ binned to reduce the noise. The position of the star is marked by an asterisk in red. The upper panel shows $Q_\varphi$ (left) and $U_\varphi$ (right) images. Lower panel: Isophotal contours of polarized light overlying $Q_\varphi$ image. The contours were measured from the $Q_\varphi$ image smoothed via a Gaussian kernel with $\sigma =$ 1.5 px. Contour levels are given for 3 (blue line), 9 (light blue), 15 (orange) and 21 (red) counts per frame per binned pixel. 
The white dotted line shows the position of the expected ring axis. 
The region inside the white stellarcentric circle with radius 
$\sim 0.12''$ is dominated by strong speckles variations. The color-bars show the counts per binned pixel.  
\label{Qphi}}
\end{figure*}

Figure\,\ref{Qphi} shows the final $Q_\varphi$ and $U_\varphi$. The $Q_\varphi$ 
image clearly reveals the nearly edge-on disk structure down 
to a projected separation of $\sim0.1''$. 
Polarized light is detected across the entire width of the 
image of $ \sim 3.6''$. The peak of the surface brightness 
appears here as a narrow stripe below the expected major axis of an inclined circular ring
(white dotted line) with a flux minimum near the position of the star.\\

By contrast, the $U_\varphi$ image contains no structural features from 
the disk. Assuming azimuthal polarization of light generated 
in single scattering processes and no multiple
scattering \citep[see][]{Canovas2015}, we do not expect to find any
astrophysical signal in the $U_\varphi$ image. Therefore, this image can 
be used for an estimation of the statistical pixel to pixel noise level 
and large-scale systematic errors in our observations.

Very close to the star, marked by a white circle with a $r\simeq0.12''$ (Figs. \ref{qu} and \ref{Qphi}), 
the data are unreliable because 
of strongly variable wings of the PSF peak. Also visible are the faint features at 
$r \gtrsim0.12''$ above and below the disk which
are negative in the $Q$ and $U$ images, and appear as positive signal
in the $Q_\varphi$ and $U_\varphi$ images. These features are much fainter
(factor < 0.1) than the disk signal and originate most likely from poorly 
corrected instrumental effects because an intrinsic signal is expected to produce 
no $U_\varphi$ signal. 

\section{Data analysis} \label{Data Analysis}

\subsection{Disk position angle}
We measured the position angle of the disk in the $Q_\varphi$-image
by the determination of the orientation of the mirror line through 
the central star perpendicular to the disk. The best position angle
was found by searching with an angle increment of 0.1$^\circ$ 
the orientation of the mirror line which produces the smallest
residuals if one side is subtracted from the other side.   
 
The results from the polarimetric and imaging data sets agree. After including ZIMPOL's True North offset of $-2^{\circ}$ we obtain the position angle of the disk axis to be 
$\theta_{\rm disk}=114.5^{\circ} \pm 0.6^{\circ}$. This value is in good agreement with PA = $114^{\circ}$ reported by \cite{Thalmann2013} for the scattered light images in H-band and with PA = $115^{\circ}$ measured by \cite{Lieman-Sifry2016} in the sub-mm range.

We define an $x-y$ disk coordinate system where the star is at the
origin, $+x$ is the coordinate along the major axis in roughly WNW-direction 
($\theta_{\rm disk}+180^\circ$), $-x$ towards ESE ($\theta_{\rm disk}$), and
$y$ perpendicular to this with the positive axis towards 
NNE (or $24.5^\circ$ EoN). The disk images in 
Figs. \ref{imaging data} and \ref{Qphi} and the plot coordinates in Figs.~\ref{Profiles} and \ref{4plots} are given in this system.    

Scattered light images of edge-on disks after classical ADI, LOCI or PCA-ADI reductions suffer from the disk flux over-subtraction particularly in the regions close to the star. The degree of flux loss depends on the shape of stellar PSF and, hence, on the observational conditions. This also applies to the total intensity image of the disk shown in Fig. \ref{imaging data}. In contrast, the intensity of the polarized light in the $Q_\varphi$ image is not strongly affected by the data reduction and better suited for the analysis of the disk structure. Therefore, in the following sections, we study, model and discuss the distribution of the polarized surface brightness based on the $Q_\varphi$ image.
   
\subsection{Polarized light brightness profiles vertical to the disk} \label{Vertical brightness profiles}
\begin{figure*}
   \centering
   \includegraphics[width=7.53cm]{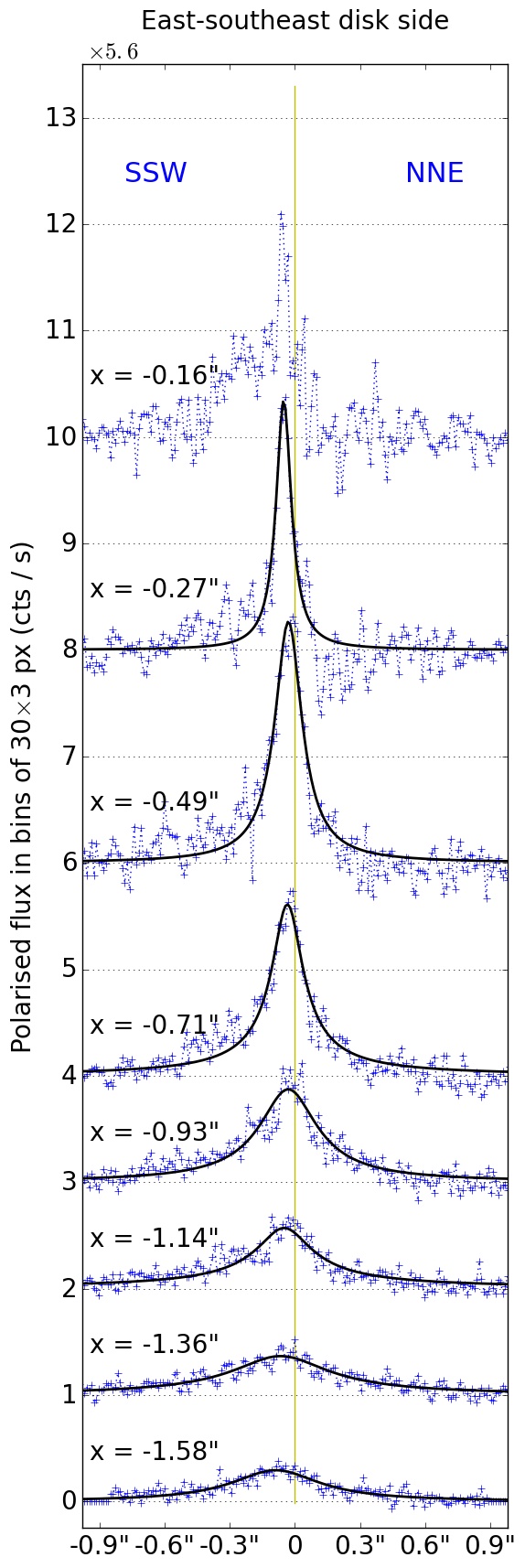} 
   \includegraphics[width=7.015cm]{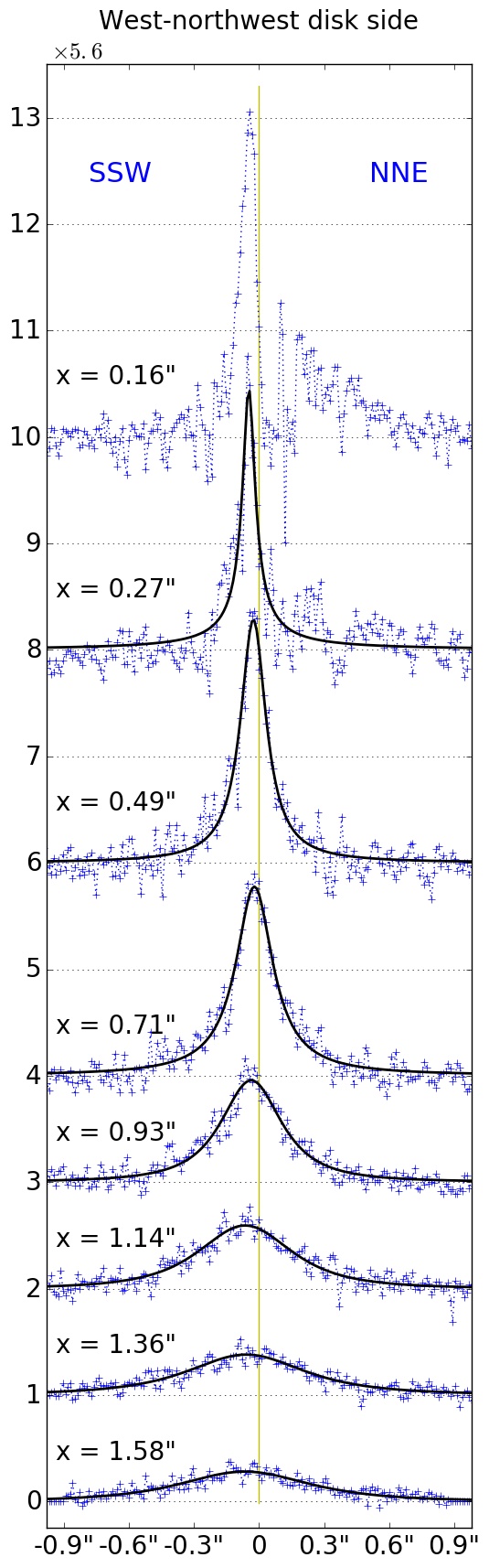}\\
   \fontsize{12}{18}\selectfont
   \text{Vertical separation from the disk major axis}\par

               \caption{HIP 79977 polarized intensity cross-sections 
perpendicular to the disk major axis at several separations $x$ from the 
central star. 
Blue crosses are the data from the $30\times3$ px binned $ Q_\varphi$ image. 
Black solid lines show the Moffat profile fits to the data except for 
$x=\pm 0.16"$, where the data are unreliable because of systematic effects.
The cross-sections are offset vertically by integer units for clarity. 
The yellow line marks the position of the disk axis.}
               \label{Profiles}
   \end{figure*}
 
\begin{figure*}{}
   \centering
   \includegraphics[width=17cm]{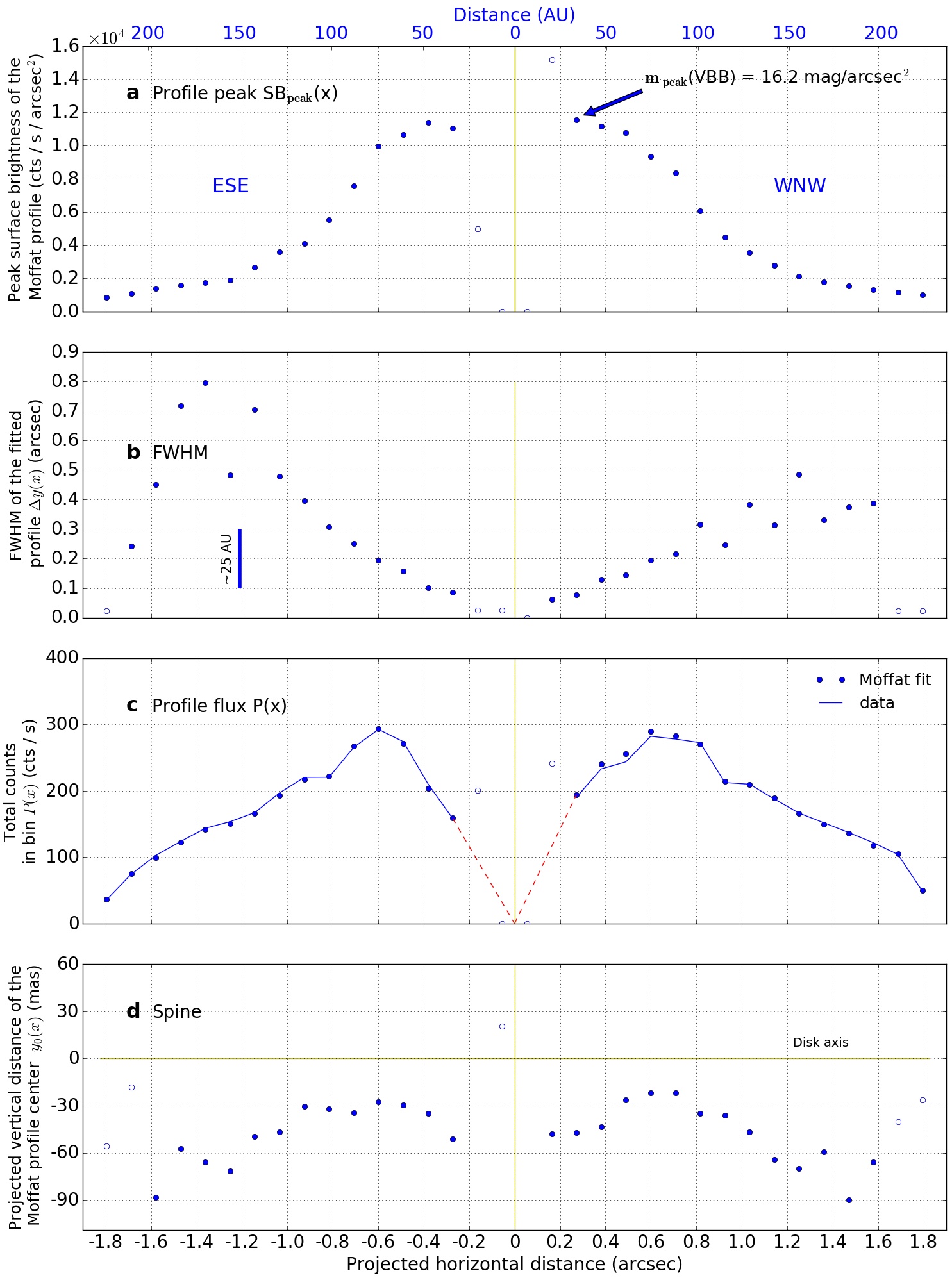}
  \caption{HIP 79977 debris disk properties along the axis $x$. The individual points give parameters of the Moffat profile 
of the vertical cross section as shown in Fig.~\ref{Profiles} 
and described in Sect. \ref{Vertical brightness profiles}. 
From the top to the bottom: 
({\bf a}) the profile peak ${\rm SB} \mathrm{_{peak}}(x)$, 
({\bf b}) FWHM, ({\bf c}) vertically integrated flux $P(x) $, 
and ({\bf d}) spine distance from the disk major axis $y_0$ 
The vertically integrated profile flux $P(x) $ is calculated as a 
mean surface brightness in a 0.1$''\times 1.8''$ bin. At separations smaller
than $x\approx0.2''$ the systematic uncertainties are increased and
open circles mark the low S/N points. The vertical yellow line indicates 
the position of the star.} 
         \label{4plots}
\end{figure*}  
                      
Figure \ref{Profiles} shows the vertical brightness profiles 
at different separations $x$ from the star which are obtained
from the $Q_\varphi$ image by applying a wide binning of 30 pixels (108 mas) 
in $x$-direction and a narrow binning of 3 pixels in $y$-direction.
Obviously, the disk structure is very similar or symmetric  
on the east-southeast (ESE) and west-northwest (WNW) sides of the disk, with
strongly peaked vertical profiles at small separations 
$x\lesssim$ 0.5" ($\lesssim$ 60 AU) and weak and broad profiles at
large separations $x\gtrsim 0.7''$ ($\gtrsim$ 87 AU). 
The innermost profiles at $x=\pm 0.16''$ and also slightly at 
$x=\pm 0.27''$ are affected by the residual instrumental features
 restricted to small $|x|$-coordinates.  

The vertical profiles can be fitted well by the Moffat 
function \citep{Trujillo2001}  
\[f_M(y)=a_M\left[ 1+ \left(\frac{y-y_0}{\alpha}\right)^2\right] ^{- \beta},\]
where $a_M$ is the flux peak located at a 
vertical distance $y_0$ from the disk major axis. 
The parameter $\alpha$ and exponent $\beta$ are related to the FWHM by 
\[\mathrm{FWHM}_M= \Delta y= 2 \alpha \sqrt{2^{\frac{1}{\beta}}-1}.\]
We used a non-linear least squares algorithm to find the best fit
parameters for the vertical Moffat profiles.

Figure~\ref{4plots} shows the $x$-dependence of the vertical profiles along
the major axis of the disk. The top panel (Fig. \ref{4plots}(a)) 
demonstrates the nearly identical decrease of the profile's peak 
as a function of the projected separation $\pm x$ for both sides of the 
disk. The profiles with the highest peak flux $a_M$ lie between $x=\pm 
(0.20''\,{\rm and}\, 0.45'')$. The results of our measurement of interior 
$r\approx0.2''$ cannot be considered as reliable because of the residual speckle noise and detector
saturation effects.\\

As shown in Fig. \ref{4plots}(b), the disk 
width $\Delta y$ is continuously increasing with separation $|x|$ from 
about $\Delta y\approx 0.08''$ ($ \sim$10 AU) at $x=0.2''$ ($ \sim$25 AU) 
to $\Delta y=0.3''$ ($ \sim$37 AU) at $x=0.8''$ ($ \sim$100 AU). At 
$|x|>1$" the disk width is not well defined but the ESE side seems to
be broader than the WNW side. The points beyond $x=1.6''$ are not 
included in this estimate because of the low SNR at large separation.\\  

The blue line in Fig. \ref{4plots}(c) gives 
the vertically integrated polarized flux $P(x)$ per $\Delta x$-interval (width 108 mas)
along the major axis. The integration in $y$-direction is from
$ y= -0.9''$ to $+0.9''$ for each x-bin. The blue dots are the same
but the integrated flux is derived from the fitted Moffat profiles.
According to this, the maximum brightness in polarized light of the 
edge-on disk in HIP 79977 is at a separation of $x=0.6''$ ($\sim$74 AU).
There is a very small discrepancy between data and fit for $x\lesssim0.6''$ because the Moffat profile cannot fit correctly negative flux
values at small angular separations which originate from the systematic 
effects described above.  

The vertical offset $y_0(x)$ of the disk spine is shown in Figure \ref{4plots}(d). The  spine curve 
is roughly symmetric with respect to $ x_0$. The smallest $y_0$-offset 
is approximately $-25\pm 5$ mas (2.5 AU) around $x\approx 0.6''$ $\pm 0.1''$. 
Closer to the star, $x\approx \pm 0.3''$, the spine is further away from
the major axis with $y_0\approx -50$~mas, and also in the outskirts 
($|x|\gtrsim 1''$) the $y_0$-offset is even more than 50~mas. In comparison, the offset $y_0(x)$ of the disk spine measured in the imaging data (Fig. \ref{imaging data}) is approximately $-60\pm 5$ mas ($\approx $7.5 AU) at $|x|<0.3''$. For larger separations, the $y_0$-offset in intensity is smaller and achieves a minimum $\approx - 45$~mas at $|x|=0.7'' \pm 0.05''$.

\subsection{Polarized flux, surface brightness and contrast} \label{Contrast}
The polarimetric image in Fig. \ref{Qphi} and the deduced profiles in Fig. 
\ref{4plots} serve as basis for the quantitative determination of the polarized
flux and surface brightness of the disk which can both
be compared to the stellar brightness with ``contrast'' parameters. 

We derive the total polarized flux of the debris disk by summing
up all the bins from $|x|=0.3''$ to 1.8$''$ along the major axis
in the integrated flux profile $P(x)$ given in Fig.~\ref{4plots}(c).
This does not include the innermost regions $|x|<0.2''$. Only
a small polarization signal is expected at small apparent separations
for a disk or ring with an inner radius $r>0.2''$, because at small separations we observe
scattering from the disk sections located in front of and behind the star. This forward
and backward scattering produces only little polarization. Thus, one
can approximate the innermost disk with a linear extrapolation 
of the measured curve from $P(x=0.27'')$ to $P(x=0.0'')=0$ 
(red dotted line in Fig.~\ref{4plots}(c)). 

This neglects a possible contribution of polarized flux from warm dust located 
very close ($r<0.2''$) to the star. Studies on the spectral energy distribution of HIP 79977 \citep[e.g.,][]{Chen2011}
indicate that there is no significant ($\gtrsim 1\%$) signal to the IR excess emission from warm dust at small separation. 
Therefore, we assume that there is also no significant unresolved contribution from an inner disk to
the polarization signal.

The polarized flux in the VBB filter, 
covering an effective aperture area of $3.6'' \times1.8''$ and including the interpolated points inside interval $|x|< 0.3''$, amounts to 5800 counts per second and per ZIMPOL arm. 
This value must be corrected for the variable atmospheric transmission $T_{\rm {atm}}$ (Fig.~\ref{counts}) using
a factor of $f_{\rm {corr}}=1/T_{\rm {atm}}= 1.3\pm 0.1$. 
This yields a corrected count rate of 7540 $\pm$ 800 cts/s where the uncertainty 
is dominated by $f_{\rm corr}$. 

The determination of the stellar flux of HIP 79977 must account for the saturation of
the PSF core and the cloudy weather. We first determine the mean value of $1.13\cdot10^7$ cts/s in the VBB filter for frames 210-280 which were apparently not affected by clouds (Fig.~\ref{counts}).   
Because the exposure is saturated out to 
the radius $r \cong 3$ px some flux is lacking. To account for the saturated part of the 
PSF, we compare the HIP 79977 profile with high-quality ZIMPOL PSFs of the standard star HD 183143 \citep[STD261\_0013-24, ][]{Schmid2017}, which were taken
under excellent atmospheric conditions. For the narrow band filters N\_R ($\lambda_c=646$ nm, $\Delta\lambda= 57$ nm) and N\_I ($\lambda_c=817$ nm, $\Delta\lambda= 81$ nm),
these PSFs contain 
within a radius of $r=5$ px a flux between $\sim20 $\% and $\sim25 $\% of the total
stellar flux measured for an aperture of $3''$ diameter. Based on this, we assume for our
HIP 79977 data, that the round annulus with inner and outer radii $r_{\rm in}=5$ px and $r_{\rm out}=416$ px 
($= 3''$ diameter) contains between $\sim75$\% and $\sim80$\% of the flux expected for an
unsaturated PSF profile. This yields for the corrected stellar count rates 
between $1.33\cdot10^7$ cts/s and $1.40\cdot10^7$ cts/s per ZIMPOL arm for observations
in the VBB filter in the slow polarimetric mode.   

The count rates are converted to photometric magnitude $m$(VBB) using the following 
expression \citep{Schmid2017}:
\[m(\mathrm{VBB}) = -2.5 \log (\mathrm{cts/s})-\mathrm{am}\cdot k_1(\mathrm{VBB})-m_{\mathrm{mode}}+ \mathit{z}p_{\mathrm{ima}}(\mathrm{VBB}) ,\]
where am~$=1.15$ is the airmass, $ k_1(\mathrm{VBB})=0.086^m$ is 
the filter coefficient for the atmospheric extinction, 
$\mathit{z}p_{\mathrm{ima}}(\mathrm{VBB})=24.61^m$ is the photometric 
zero point for the VBB filter and $m_{\mathrm{mode}}=-1.93^m$ is an offset 
to the zero point which accounts for the used instrument and detector mode. 
We obtain for HIP 79977 a magnitude $m(\mathrm{VBB})$ = 8.60$^m \pm 0.07^m $ 
in good agreement with the literature values (see Table \ref{t_filters}).
The derived photometric magnitude $m(\mathrm{VBB})$ yields the color V-VBB = 9.09$^m $ - 8.60$^m $ = 0.49$^m $ which is close to the color index in the Johnson-Cousins' photometric system $\mathrm{V}-\mathrm{I}_C = 0.44^m $ \citep[$\lambda_{\rm eff}$ = 0.806 $\mu$m,  $\Delta\lambda= 0.154$ $\mu$m for $\mathrm{I}_C$; ][]{Pecaut2012} for a F2/3V star.
\begin{table}  
      \caption[]{HIP 79977 photometry.  \label{t_filters} }

		  \centering
	     \begin{tabular}{lccccc}
            \hline
            \hline
            \noalign{\smallskip}
            Filter & $\lambda$ & $\Delta \lambda$ & mag & $\sigma_{\rm mag}$ & Ref. \\
                   & ($\mu$m) & ($\mu$m) & (mag) &  (mag) &  \\
            \hline
            \noalign{\smallskip}
	   	    HIP H$_{\rm P}$ & 0.528 & 0.221 & 9.20 & <0.01 & 1 \\ 
            Tycho V & 0.532 & 0.095  & 9.11 & 0.02 & 2 \\
            Johnson V & 0.554 & 0.082 & 9.09 & <0.01 & 1 \\ 	
            Gaia G  & 0.673 & 0.440 & 8.93 & <0.01 &  3\\
            {\bf ZIMPOL VBB}  & 0.735 & 0.290 & 8.60 & 0.07 &4  \\           
            Johnson J & 1.250 & 0.300 & 8.06 & 0.02 &5  \\

	   \noalign{\smallskip}
            \hline
            \noalign{\smallskip}
         \end{tabular}\\
           \begin{minipage}{8.5cm}%
    \tiny {\bf Notes}: (1) \cite{ESA1997}; (2) \cite{Hog2000}; (3) \cite{GaiaCollaboration2016}; (4) this work; (5) \cite{Cutri2003}.%
  \end{minipage}
  \\
   \end{table}

For the polarized flux of the whole disk we get 
$mp_{\mathrm{disk}}(\mathrm{VBB})$ = 16.6$^m$ $\pm$ 0.3$^m$. This yields a ratio 
of total polarized flux of the disk to the stellar flux of $(F_{\rm pol})_{\rm disk}/F_{\rm \ast}=
(5.5 \pm 0.9)\cdot 10^{-4}$. \\

We determine for the peak surface brightness of the polarized light 
${\rm SB} \mathrm{_{peak}(VBB)} = 16.2^m$ arcsec$^{-2}$ along the inner 
($0.2''-0.4''$) disk spine (Fig. \ref{4plots}(a)) and 
a surface brightness contrast for the polarized flux of 
${\rm SB} \mathrm{_{peak}(VBB)} - m \mathrm{_{star}(VBB)} = 7.64$ mag arcsec$^{-2}$.  
For the outer disk around $x\approx \pm 1.7''$ the surface brightness contrast
is about 10~mag arcsec$^{-2}$.

\begin{figure}  
   \centering
   \includegraphics[width=9cm]{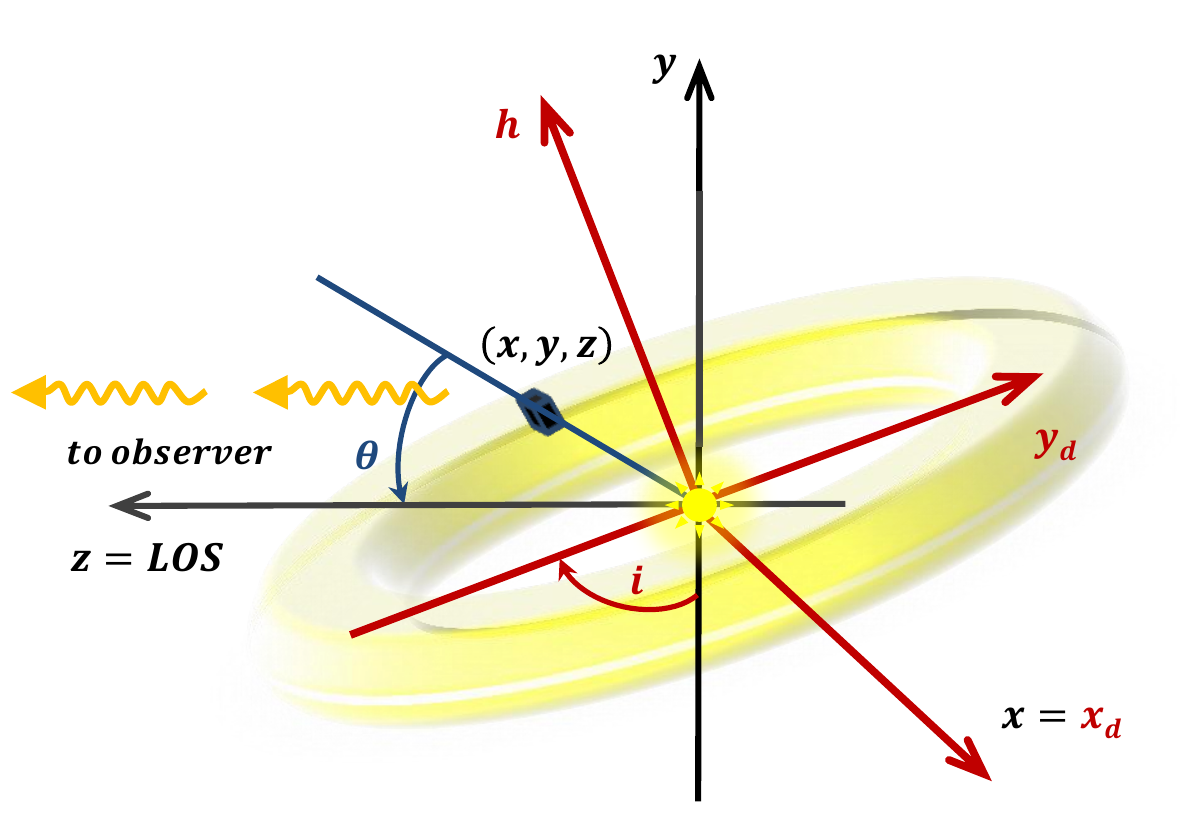}

               \caption{Illustrative sketch of the debris disk with inclination \textit{i} and coordinate systems $(x, y, z)$ and $(x_d, y_d, h)$ used in model. The small blue cube at scattering angle \textit{$\theta$} marks the position \textit{(x, y, z)} of a grid element with grain number density \textit{n(x, y, z)}.  \label{Disk_skizze}}
 \end{figure}

\section{Modeling} \label{Modeling}
To reproduce the physical appearance of the debris disk around 
HIP 79977 we construct a 3D model 
for the scattered intensity and the polarization flux from optically thin (single scattering) dust. 
The disk is described by an axisymmetric dust distribution 
using the cylindrical coordinates $r=\sqrt{x^2_d+y^2_d} $ and \textit{h}, 
where $x_d$ and $y_d$ 
describe the disk midplane and the axis $h$ gives the height above it 
(see Fig.~\ref{Disk_skizze}). The disk 
model is projected onto an x-y sky plane, 
where $x=x_d$ defines the line of nodes and y is the
perpendicular axis through the central star. The $z$-axis is 
equivalent to the line of sight to the star and the z-component is important for 
the calculation of the scattering
angle $\theta$. The disk coordinates are related to the sky coordinates by: \\
$x = x_d$\,, \\
$y = y_d \cos i + h \sin i\,,$ \\
$z = -y_d \sin i + h \cos i \,.$ \\

Following \cite{Artymowicz1989} we adopt a product of two functions to describe the number density distribution \textit{n(r, h)} of dust grains in the disk
\[n(r, h) \sim R(r)\ Z(h).\]
For the radial \textit{R(r)} and vertical \textit{Z(h)} distribution profiles we adopt expressions which are often used in the literature \citep{Augereau2001, Ahmic2009, Thalmann2013} in accordance with the theory of a "birth ring", a planetesimal reservoir in analogy to the Kuiper Belt in the solar system. In this ring, dust down to sub-micron sizes is produced by collisions and evaporation of solid bodies. The radial profile is given by the following expression:
\begin{equation} \label{R}
R(r) = {\left( \left(\frac{r}{r_0}\right)^{-2\alpha_{in}}+\\\ \left(\frac{r}{r_0}\right)^{-2\alpha_{out}} \right)}^{-1/2},
\end{equation}
where $r_0$ is the radius of planetesimal belt and radial power laws $r^{\alpha_{in}}$ $(\alpha_{in}>0)$ and $r^{\alpha_{out}}$ $(\alpha_{out}<0)$ describe the increase of grain number density inside the "birth ring" and the decrease of the density in the outer region, respectively. The vertical profile \textit{Z(h)} defines an exponential drop-off with the disk height:
\begin{equation} \label{Z}
Z(h)= \exp\left[- \,{\left(\frac{|h|}{H(r)}\right)}^\gamma\right],
\end{equation}
where $\gamma =1$ for a purely exponential fall off and $\gamma =2$ for the Gaussian profile. For the scale height $H(r)$ we assume a power law dependence on radius
\[H(r)= H(r_0)\,{\left(\frac{r}{r_0}\right)}^\beta,\]
where $H(r_0)$ is a scale height at $r_0$ and $ \beta$ is the flare index of the disk.

For an optically thin debris disk the amount of scattered radiation 
from a volume element with coordinates $(r, h)$ is determined by the intensity 
of the incident light at wavelength $\lambda$ and the product of the average grain cross-section 
for scattering $\langle \sigma_{sca}\rangle (r,h)$ per particle with the number 
density $n(r,h)$ of grains in this volume. How much light is scattered by particles into 
the specific direction depends on the scattering 
angle $\theta$:
\[\theta = \arccos \left( \frac{z}{\sqrt{x^2+y^2+z^2}}\right)\]
and is described by the phase function $f_{\lambda}(\theta)$. We derive the intensity of the 
light in the computed image from the integral over all grid cells along the line 
of sight or z-axis
\begin{equation} \label{eq:In}
I_{\lambda}(x, y)=\frac{L_{\lambda}}{4 \pi D^2}\int\frac{f_{\lambda}(\theta) \, \langle \sigma_{sca, \, \lambda}\rangle (r,h)\, n(r,h)}{4 \pi(x^2+y^2+z^2)}dz, 
\end{equation}
where $ L_{\lambda}$ denotes the HIP 79977 monochromatic luminosity at wavelength ${\lambda}$, $D$ is the star-Earth distance and $f_{\lambda}(\theta)$ is an averaged 
dust scattering phase function (see Sect. \ref{phase f}).\\ 
The grain cross-section for scattering $\sigma_{sca, \, \lambda}$ is a product of the grain geometrical cross-section with the grain-scattering efficiency $Q_{sca}$.
In general, the scattering efficiency as well as the phase function depend 
on the wavelength of the incident light $\lambda$ and the grain size, shape and composition. Assuming the same composition and shape parameters for all grains in the unit volume with coordinates $(r, h)$, we can average over all particle sizes to express $\sigma_{sca, \,  \lambda}(r, h)$ per particle as 
\begin{equation} \label{eq:sigma}
\begin{split}
\langle \sigma_{sca, \,  \lambda} \rangle (r, h) = \pi \ \langle Q_{sca, \,  \lambda}(a)\, a^2\rangle = \\ 
= \frac{\pi}{n(r,h)}\int\limits_{a_{min}}^{a_{max}} Q_{sca, \,  \lambda}(a) \, a^2\,n(a) da,
\end{split}
\end{equation}
where $a$ is a grain radius varying between the minimum size $a_{min}$ and maximum size $a_{max}$ for a given grain size distribution $n(a)$, and $n(a)\, da$ defines the differential number density of grains with radii in the interval $[a, a+da]$. The grain minimum and maximum sizes have to be fixed in our model if the phase function is calculated from the Mie scattering theory. In detailed treatments these parameters can vary freely but in order to reduce the running time of the code, we simplify the computation of the scattering cross-section by considering the same grain-size distribution, grain sizes and optical properties everywhere in the disk. In this case the average cross-section per particle is constant through the disk and we can take it out of an integral: 
\begin{equation} \label{eq:In_2}
\begin{split}
I_{\lambda}(x, y)=\frac{L_{\lambda} \langle \sigma_{sca, \, \lambda} \rangle }{4 \pi D^2}\int\frac{f_{\lambda}(\theta)\, n(r,h)}{4 \pi(x^2+y^2+z^2)}dz = \\
= A \int\frac{f_{\lambda}(\theta)\, R(r)\ Z(h)}{(x^2+y^2+z^2)} dz,
\end{split}
\end{equation}
where $A$ is a normalization parameter containing all constants used in the model, such as the HIP 79977 luminosity and the star-Earth distance, and so on.  \\

In this work we concentrate on the polarized scattered light from the debris disk. 
Therefore we need to model the polarized flux, which requires the consideration
of a different scattering phase function $f_{\lambda}(\theta,g_{\rm sca})$ together with the corresponding
angle dependence of the produced polarization signal 
$p_m(\lambda)LP(\theta)$ as discussed in the following subsection. The result
follows then from the integration
\begin{equation} \label{eq:In_3}
\begin{split}
P_{\lambda}(x, y)=\frac{L_{\lambda} \langle \sigma_{sca, \, \lambda} \rangle }{4 \pi D^2}\int\frac{p_m(\lambda) \, LP(\theta) \, f_{\lambda}(\theta,g_{\rm sca})\, n(r,h)}{4 \pi(x^2+y^2+z^2)}dz = \\
= A_p \int\frac{LP(\theta) \,  f_{\lambda}(\theta,g_{\rm sca}) \, R(r)\ Z(h)}{(x^2+y^2+z^2)} dz,
\end{split}
\end{equation}
where $A_p$ is the scaling factor $A\cdot p_m$.

The model images for the different polarization components $I_0$, $I_{90}$,
$I_{45}$ and $I_{135}$ must be convolved with an instrument PSF before being combined to the model images of the Stokes parameters which can be compared with the observations. Because the PSF shape 
is strongly variable, we
selected a mean PSF which is representative for the observations. This mean PSF was fitted
with a radial, rotationally symmetric Moffat profile which was used for the convolution. 
The exact shape of the stellar PSF is not so critical because our disk models have a
relatively simple structure.

\subsection{The scattering phase function for polarized light} \label{phase f}
The phase function (PF) $f_{\lambda}(\theta)$ in Equation~\eqref{eq:In} characterizes the angle 
dependence of scattered radiation. In the following, we disregard the wavelength dependence of the PF.

A very popular way to describe the scattering phase function is 
the Henyey-Greenstein (HG) function \citep*{Henyey1941}:
\begin{equation} \label{eq:g}
f(\theta)=\frac{1-g^2}{4 \pi (1+g^2-2 g cos \theta)^{3/2}},
\end{equation}
where $g$ is the average of the cosine of the scattering angle which characterizes the shape of the phase function. For isotropic scattering $g=0$, forward scattering grains have $0<g\leq 1$, while for $-1\leq g<0$ the scattering is peaked backwards. 

However, there exists also growing evidence that a simple HG-function is a poor 
approximation for the modeling of the scattered intensity from debris disks. 
This is nicely demonstrated for the bright disk HR 4796A \citep{Milli2017}, which shows,
for small phase angles $\theta<30^\circ$, a strong diffraction peak and, for large phase 
angles $\theta>30^\circ$, a scattering intensity which is roughly angle-independent. Thus,
a more general phase function, for example, a two-component (or double) HG function seems to be 
required for the modeling of the scattered intensity of highly inclined debris disks 
  
\begin{equation} \label{eq:phase func}
f(\theta,g_{\rm diff},g_{\rm sca})=w \cdot f(\theta,g_{\rm diff}) 
+ (1-w) \cdot f(\theta,g_{\rm sca})\,,
\end{equation} 
where the first term describes the strong diffraction peak,
the second term represents the more isotropic and much less forward scattering part \citep[see also][]{Min2010}, and $w$ is the scaling parameter, $0\leq w\leq 1$.

For the polarized scattered radiation from a debris disk the
situation is slightly different. The strong forward peak seen in intensity, which
can be ascribed to the light diffraction by large particles $a\gg \lambda$, 
is expected to produce no significant 
light polarization. The scattering polarization is produced by the photons hitting the
particle surface and interacting by diffuse reflection or/and refraction and transmission
as described above by the second term $f(\theta,g_{\rm sca})$. But, in addition, the
angle dependence of the linear polarization $LP(\theta)$ produced by the particle scattering needs
to be taken into account. For example, strict forward and backward scattering will
produce no polarization for randomly oriented particles for symmetry reasons. We adopt the Rayleigh scattering function as a simple approximation for
the angle dependence of the polarization fraction $p_{\rm sca}$:

\[p_{\rm sca}(\theta) = p_m\frac{1- \cos^2\theta}{1+\cos^2 \theta}= p_m LP(\theta),\]
with the scaling factor $p_m$, which defines the maximum fractional polarization produced
at a scattering angle of $\theta=90^\circ$.

\begin{figure}  
   \centering
   \includegraphics[width=8cm]{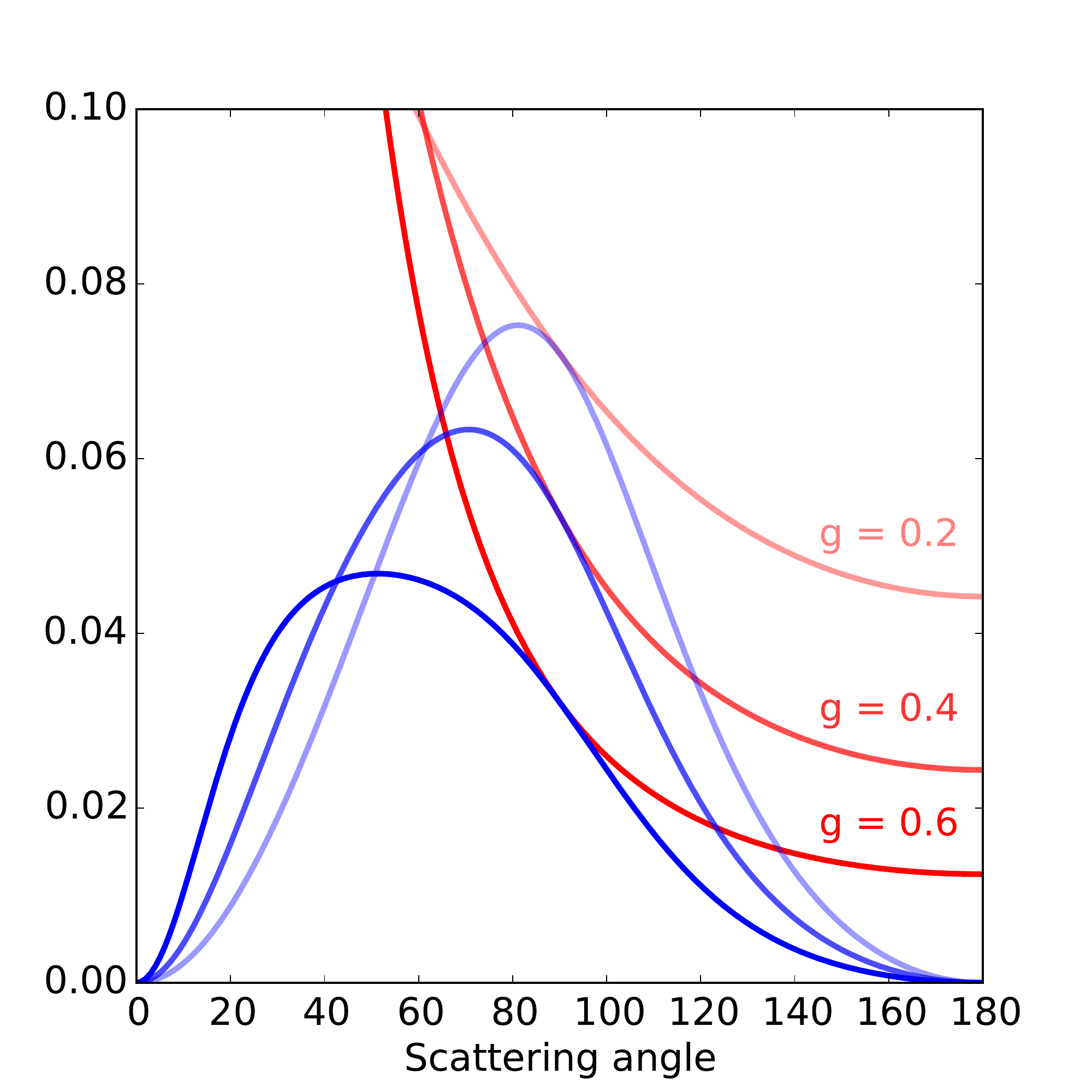} 

               \caption{Scattering phase function for the polarized light (blue)
for three different asymmetry parameters $g_{\rm sca}=0.2$, $g_{\rm sca}=0.4$, 
and $g_{\rm sca}=0.6$. Red lines show the corresponding Heyney-Greenstein 
functions for $f(\theta,g_{\rm sca})$.  \label{Phase_functions}}
   \end{figure}

Figure \ref{Phase_functions} shows some examples of obtained phase
function for the polarized flux $ LP(\theta)f(\theta,g_{\rm sca})$ for different
cases of the HG function $f(\theta,g_{\rm sca})$. For isotropic 
scattering ($g_{\rm sca}=0$) the maximum of scattered polarized flux occurs
at $\theta = 90^{\circ}$. For an asymmetry parameter $g_{\rm sca}>0$ the maximum 
is shifted to smaller scattering angles producing a corresponding
asymmetry in the amount of polarized light received from the front and 
back sides of the disk. So, for example, the value of polarized 
flux PF ($g_{\rm sca}=0.6$) at $\theta = 20^{\circ}$ is 35 times higher 
than at $\theta = 160^{\circ}$.

\subsection{Model fitting} \label{Fitting}
We have calculated $5.28\cdot10^6$ models for a parameter grid as specified in Table \ref{Grid}
in order to find the set of model parameters which best fit the observed polarized intensity image. 

\begin{figure}  
   \centering
   \includegraphics[width=9cm]{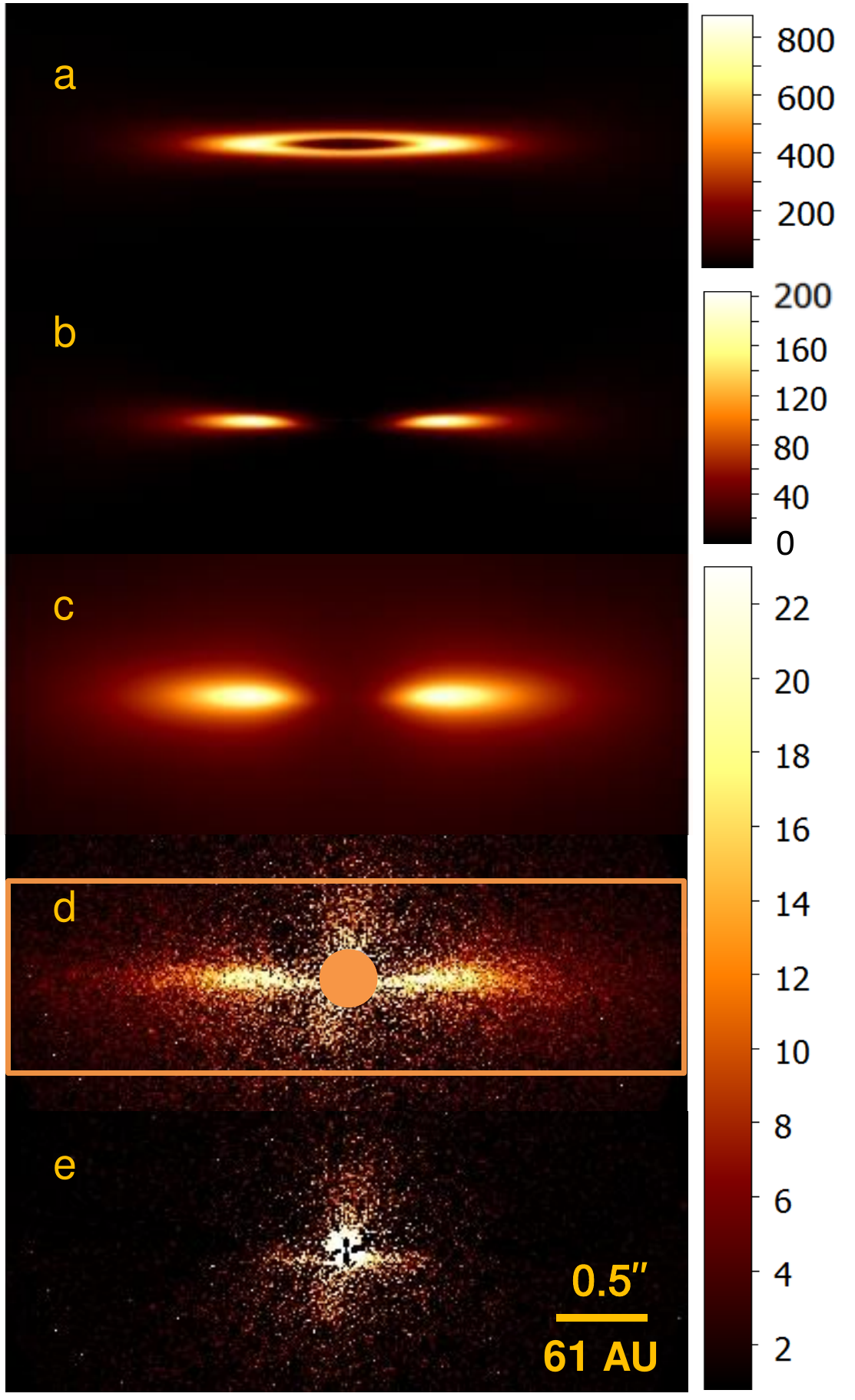} 

               \caption{Comparison of the best-fit model with the $Q_\varphi$ image. 
(\textbf{a}) Image visualizing the dust distribution in the disk. (\textbf{b}) Model image of the polarized light non-convolved with PSF. (\textbf{c}) Model image of the polarized light convolved with the instrumental PSF. (\textbf{d}) $Q_\varphi$ image from the data. The rectangular area outlined with an orange box shows the minimization window as described in the body text. The orange circle marks the central region of the image excluded from the $\chi^2$ evaluation. (\textbf{e}) Residual image obtained after subtraction of the PSF-convolved model image ({\bf c}) from the $Q_\varphi$ image ({\bf d}). Color-scales of images ({\bf a}) and ({\bf b}) are given in arbitrary units. The color-bar for images {\bf (c), (d)} and {\bf (e)} shows polarized flux in counts per binned pixel.  \label{model}}
   \end{figure}

For the fitting, we reduced the number of image pixels by $3\times 3$ binning and selected
a rectangular image area with a length of 341 and width of 100 binned pixels centered
and aligned to the disk x and y (major and minor) axes (see Fig. \ref{model}(d)). A round area with a radius of 16 pixels  (0.17$''$) centered on the star and the spurious features near the saturated region are
excluded from the evaluation of the fit goodness. Figure \ref{model} illustrates the 
different steps in the image fitting procedure. From the model 
dust distribution in the disk (a) the expected polarization flux is calculated (b), 
convolved with the instrument PSF (c), fitted to observation (d), and the residuals
(e) are then used for the $\chi^2_{\rm image}$ evaluation of the image fit.

The goodness of the fit was estimated for each model with the reduced $\chi^2$-parameter:
\[\chi^2_{red} = \frac{1}{N_{data}-N_{par}}\sum_{i=1}^{N_{data}} \frac{\left[ y_{i}-x_{i}(\vec{p})\right]^2 }{ \sigma_{yi}^2}, \]
where $N_{data}$ is a number of data points with measurement results $y_{i}$ which have uncertainties $\sigma_{yi}$. 
Each data point corresponds to a binned pixel within the minimization window shown in Fig. \ref{model}(d). $N_{par}$ denotes 
the number of free parameters $\vec{p}=(p_{1}, p_{2}, ..., p_{N_{par}})$ used to create a model image with 
values $x_{i}$ and listed in Col. 1 of Table \ref{Grid}.\\

\begin{figure}  
   \centering
   \includegraphics[width=9cm]{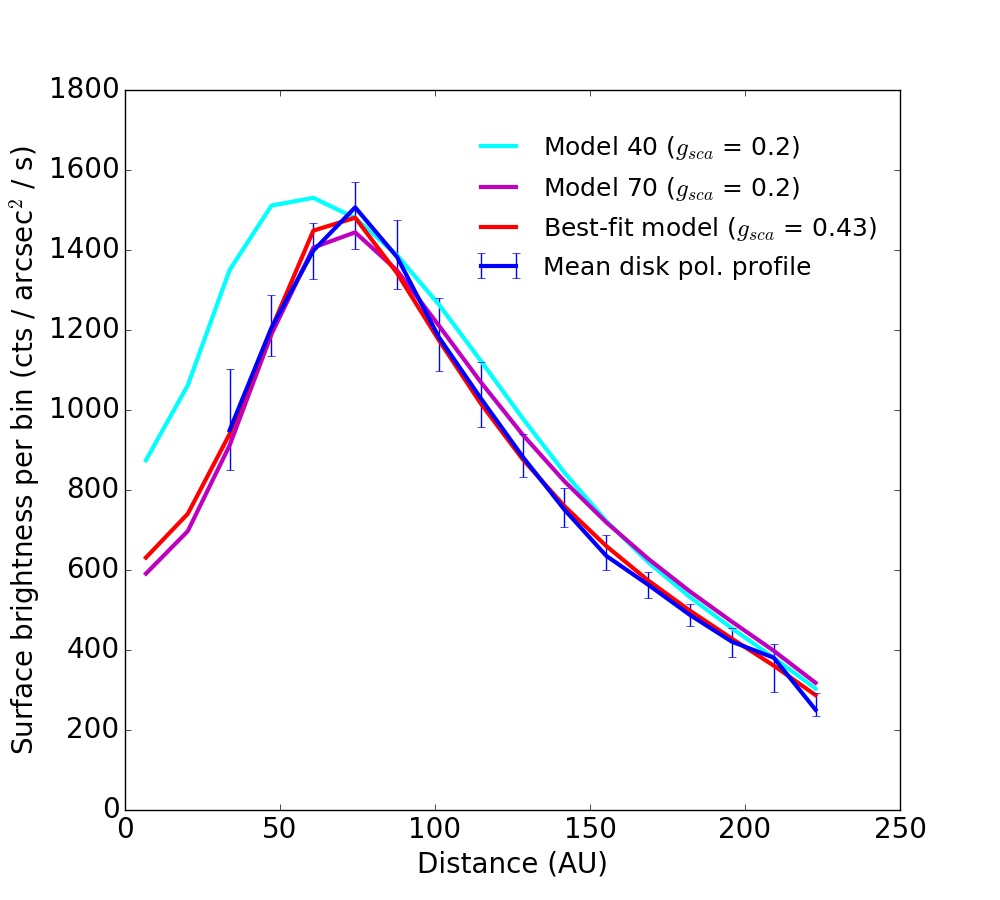}  
         \caption{Comparison of the mean disk profile $\langle P|x|\rangle$ for the polarized flux (see Sect. \ref{Fitting}) with profiles of 3 models given in Table \ref{Grid}. $\langle P|x|\rangle$ is the mean of both disk sides profiles $P(x)$ shown in Fig. \ref{4plots}(c) between 0.22$''$ (27 AU) and 1.80$''$ (220 AU). The best-fit model and "Model 70" ($\chi^2_{\rm SB}< 2.5$) fit $\langle P|x|\rangle$ while "Model 40" ($\chi^2_{\rm SB}> 2.5$) is significantly off at small distances.
\label{SBprofile}}
 \end{figure}

To accelerate the fitting procedure we have made a preselection of disk models using the mean disk profile $\langle P|x|\rangle$
along the major axis shown in Fig.~\ref{SBprofile}. The mean profile  $\langle P|x|\rangle$  consisting of 15 points from $|x| = 0.22''$ to $|x| = 1.80''$ for the observed disk polarization 
is obtained by averaging the $P(x)$ data points from the negative and positive  
$x$-axes given in Fig. \ref{4plots}(c). Thus the 2D models were collapsed to a profile and fitted first to
the $\langle P|x|\rangle$ profile calculating the $\chi^2$ and defining a good fit threshold based on the number of degrees of freedom for the fit \citep{Press2007}. 

The procedure is straight forward because the noise  
is well defined for these data points which represent flux integrations over a large area. 
This can also be inferred from the observed profiles for the two disk sides, which look essentially
identical, indicating that there are no localized spurious effects or
strong intrinsic asymmetries in the disk. The fitting does not depend on
uncertainties in the PSF model convolution because the spatial resolution is 
low. Still, the key properties of the geometric distribution of the polarized flux along the 
disk spine are captured by the $\langle P |x|\rangle$-profile. 

Models with a $\chi^2_{\rm SB}< 2.5$ are considered to fit the $\langle P|x|\rangle$-profile well (see the examples
in Fig.~\ref{SBprofile}). The profile fitting is compatible with a disk with a radius $r_0$ in the range [60, 86] AU 
which coincides with the separation of the maximum. Of course, the fitting of disk models described by 9 parameters to a 15 point $\langle P|x|\rangle$ profile cannot define a 
unique solution for HIP 79977 disk but provides more or less well defined ranges 
for the model parameters. 

The scaling factor $A_p$ (see Table \ref{Grid}) is determined by the $\chi^2$ minimization of the $\langle P|x|\rangle$-profile fit for each model. This approach has been chosen because the statistical noise is larger and not well known systematic uncertainties are much harder to quantify for the image data points.  

In a second step, we compare the 2D disk models which were preselected by the previous profile fitting to the $Q_\varphi$ image (Fig.~ \ref{model}(d)) to further constrain the
model parameters. This provides a multidimensional parameter distribution of well-fitting models 
by setting a threshold for the 2D image fit $\chi^2_{\rm image}< 8$. The mean values of the obtained 
distribution are adopted as the best-fit model parameters. Their uncertainties are given by the 68\% 
marginalized errors as calculated from the sample covariance matrix. The mean parameters together 
with the confidence intervals are listed in Table \ref{Grid} (Col. 5 and Col. 6, respectively).
The corresponding synthetic image of polarized light is shown in Fig. \ref{model}(b)
and the convolved image (Fig. \ref{model}(c)) appears to fit the $Q_\varphi$ image (Fig. \ref{model}(d)) well.
The residuals image (Fig. \ref{model}(e)) displays some PSF-shaped leftovers, the instrumental features above 
and below the disk center and, possibly, some minor residues from the disk flux. 
In this case the model would lack flux along the spine at small separation.  

\begin{table*}  
      \caption[]{Grid of parameters for the $5.28\cdot10^6$ models and resulting parameters for the best fit model. Also given are the parameters of two selected comparison models ("Model 70" and "Model 40").  \label{Grid} }
		  \centering
	     \begin{tabular}{lcccccc}
            \hline
            \hline
            \noalign{\smallskip}
            Parameter & Range & Step of linear & \multispan{2}{\hfil Best model \hfil} & Model 70 & Model 40\\
             &  & sampling & mean value & 68\% CL \\
            \hline
            \noalign{\smallskip}
            Radius of belt $r_0 \ $  (AU)& [30, 90] & 10  & 73 & 16 &70 & 40$^*$\\[3pt]
            Inner radial index $\alpha_{in}$ & [1, 10] & 1 & 5.0 & 2.8 & 2.0$^*$ & 2.0$^*$  \\[3pt]
	   	    Outer radial index $\alpha_{out}$ & [-6, -1] & 0.5 & -2.5 & 1.4 & -3.0& -2.5 \\[3pt]
	   	    Scale height $H_0$ (AU)& [0.5, 3.5] & 0.5 & 2.3 & 0.7 & 1.5 & 0.5$^*$ \\ [3pt]
	   	    Vertical profile $\gamma$&[0.5, 2.5] & 0.5 & 0.9 & 0.6 & 1.0 & 1.0\\[3pt]
	        Flare index $\beta$ & [0.5, 4.5] & 1 & 2.2 & 1.4 & 2.5 & 3.5 \\[3pt]
	        Inclination $i \ (^{\circ})$ & [82, 87] & 1 & 84.6 & 1.7 & 85.0& 82.0$^*$\\[3pt]
	        HG parameter $g_{sca}$ & [0.0, 0.9] & 0.1 & 0.43 & 0.25 & 0.20 & 0.20\\[3pt]
	        Scaling factor $A_p$ & - & - & 9.04 & - & 4.03 & 3.10\\
	   
	   \noalign{\smallskip}
            \hline
            \hline
            \noalign{\smallskip}
         \end{tabular}\\
\begin{flushleft}
Notes: $^*$ Parameter value lies outside the 68\% confidence interval.\\
\end{flushleft}
   \end{table*}

Our modeling assumes that the optical depth in the disk is small. According to our best-fit model we estimate a $\tau \approx 0.5 $ for a radial photon path through the disk midplane ($\Theta = 0^\circ$), and significantly less for $\Theta > 1^\circ$. After scattering, a photon escapes without further interaction because we see the disk inclined by $\approx 5^\circ$ with respect to edge on.

Our statistical analysis of the model fitting allows an assessment of the
parameter degeneracy problem where many different 
combinations of parameters match the data. 
In particular we notice an important degeneracy between the radius 
of the planetesimal belt $r_{0}$ and scattering asymmetry parameter $g_{\rm sca}$. 
Figure \ref{CL} shows the 68\% and 95\% confidence level (CL) regions derived from the  
distribution of these two parameters. The contours cover an 
extended region implying that the degeneracy between the radius 
of the planetesimal belt and asymmetry parameter cannot be resolved with our data. 

\begin{figure*} 
   \centering
   \includegraphics[width=16cm]{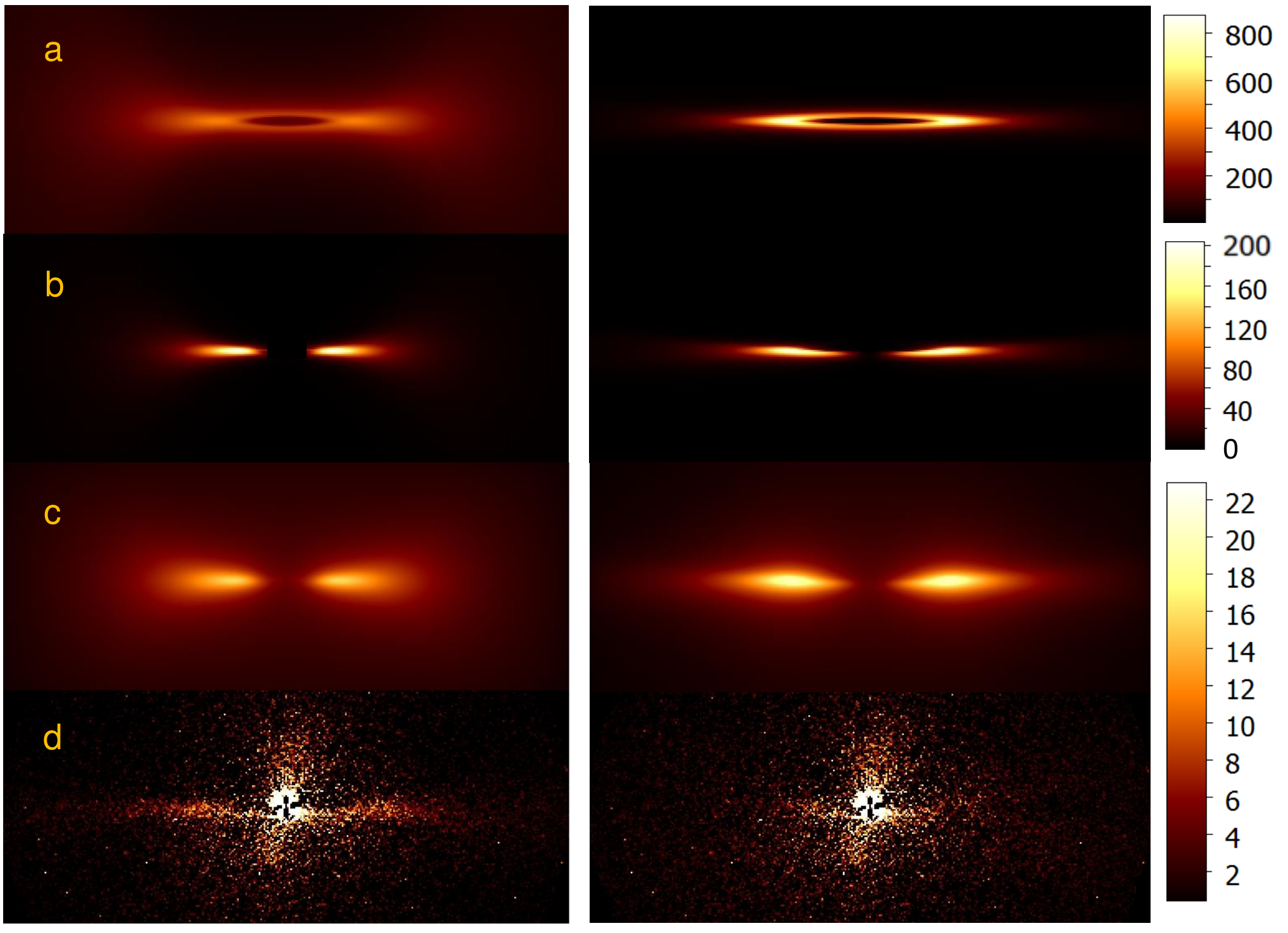}
               \caption{Comparison between alternative models of the HIP79977 debris disk: "Model 40" with the radius of the planetesimal belt $r_0=40$ AU (left) and "Model 70" with the radius of $r_0=70$ AU (right) and fitting parameters as specified in Table \ref{Grid}. From top to the bottom: ({\bf a}) model of the dust distribution in the disk, ({\bf b}) model of the polarized light, ({\bf c}) model of the polarized light convolved with the instrumental PSF and ({\bf d}) residuals left after subtraction of the PSF-convolved model image from the $Q_\varphi$ image. Color-scales of images ({\bf a}) and ({\bf b}) are given in arbitrary units. Color-bar of images ({\bf c}) and ({\bf d}) shows flux in counts at each pixel of the image.  \label{2models}}
\end{figure*}

\begin{figure}  
   \centering
   \includegraphics[width=9cm]{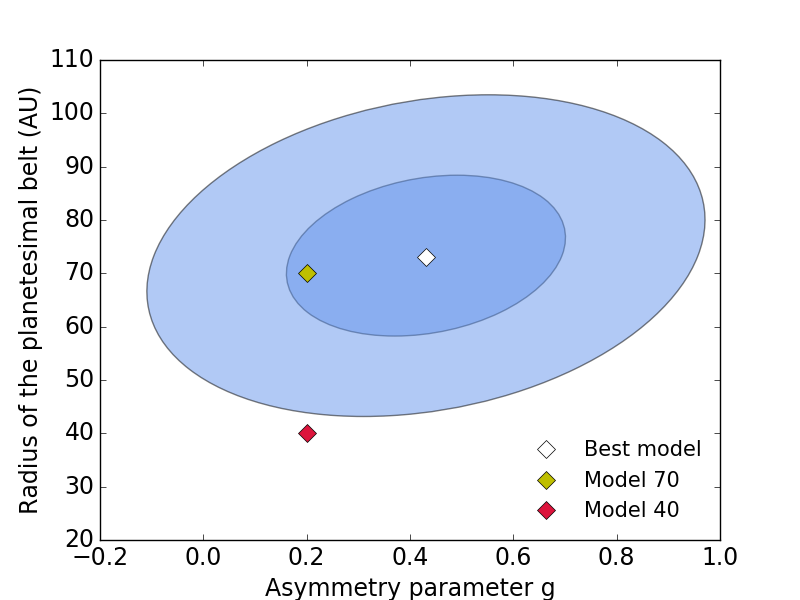}
               \caption{Two-dimensional constraints on the radius of the planetesimal belt $r_{0}$ and the asymmetry parameter $g_{sca}$ for HIP 79977. The contours show the 68\% and 95\% CL regions for the model sample and the dots point out the locations of the described models in this parameter plain.  \label{CL}}
 \end{figure}

To examine how well/badly models other than the mean model reproduce the data, we compare two models 
randomly picked from the generated distribution: one model (specified in Table \ref{Grid} 
as "Model 70") with all parameters lying inside of the 1-$\sigma$ area with the 
belt radius $r_0=70$ AU close to the mean value of this parameter, and one model 
(specified in Table \ref{Grid} as "Model 40") with the same $g_{\rm sca}$ but 
$r_0=40$ AU lying outside of the 1-$\sigma$ range. Figure \ref{2models} shows 
both models in four different views: dust distribution in the disk 
$n(y, z)$, non-convolved model image of the polarized flux, polarized image produced after the 
combination of convolved intensities $I_0$, $I_{90}$, $I_{45}$, $I_{135}$.
with the instrumental PSF. 

"Model 40" gives a significantly worse fit for the central part of 
the $Q_\varphi$ image compared to "Model 70" based on the derived $\chi^2$ and 
visual examination of the residues. The comparison of the disk polarization profile of
``Model 40'' with the observations also shows a relatively poor match (see Fig.~\ref{SBprofile}).
"Model 70" gives a reasonable fit to the polarization profile and also the residuals in the 2D image appear to be not much larger than the best-fit model, as is expected for a model within the 1$\sigma$ confidence area.

\section{Discussion} \label{Discussion}
\subsection{Disk structure}
Our results from the modeling of the dust distribution around HIP 79977 
indicate a mean radius of $ \sim$73 AU for the planetesimal belt. 
The vertical distribution of the dust in the disk is described by a profile 
with an exponent $\gamma$ smaller than two. This is a steeper fall-off 
than a Gaussian distribution, indicating a higher concentration of 
particles in the midplane. The radial distribution of the grain number density matches the shape 
of an annular disk with an inner cavity. This assumption is supported by the 
SED of HIP 79977 showing no significant thermal emission at wavelengths $\lesssim 14\, \mu$m. 
The depletion of scattering material inside a possible belt of parent planetesimals can be caused 
by the radiation pressure or drag forces acting on small particles \citep[][and references therein]{Wyatt2008}.

The radiation pressure pushing outward the dust grains with sizes close to
or smaller than the blow-out size (< 1 $\mu$m) could be responsible for 
the growing width of the disk vertical cross-sections when the separation 
from the star increases from $0.2''$ to $>1''$ (Fig. \ref{4plots}(b)). 

In the past years several authors have derived a distance of the 
dust grains from the star in HIP 79977 by modeling the shape of the 
disk SED with a single- or double-temperature fit. 
Assuming that the dust grains emit radiation like black bodies, 
\cite{Chen2011} have determined a dust grain temperature of 89 K based 
on the \textit{Spitzer} MIPS photometry. They considered amorphous silicates with 
olivine composition as the main component of the dust, and the average size 
of the grains, which were not removed by the radiation pressure, 
to be 1.5 $\mu $m. They have estimated that if the grains are spherical 
and in radiative equilibrium they should be located at a distance of at 
least 40 AU from the star. In reality, the bulk of the dust could have a larger radial separation because the real dust grains emit radiation less efficiently than black bodies. Dust with the same equilibrium temperature can therefore exist at larger distances from the star. This supports our mean model indicating a separation which is more like 70~AU.

\cite{Jang-Condell2015} postulated a much larger average grain size of
11.1~$\mu$m based on an analysis including \textit{Spitzer} IRS spectra. 
They derived a grain temperature of 102~K (for amorphous silicates with 
olivine composition) requiring 
a stellocentric distance of about 11.5 AU for the grain distribution
which is in conflict with our results. 

Previous imaging and polarimetric imaging of the disk 
around HIP 79977 in the H-band was presented by \cite{Thalmann2013}.
From the data they derived a disk orientation of 114$^\circ$ (major
axis), and an inclination of 84$^\circ$ in very good agreement with
this work. \cite{Thalmann2013} modelled the self-subtraction
effects for flux extraction for the intensity image and derived
the intrinsic intensity slope for the disk along the major axis. 
In addition, they detected a polarimetric signal 
from the disk for separations from $0.3''$ to $1.5''$,
compared the polarization with the intensity profile 
and found a fractional polarization of
$\sim 10~\%$ ($1\sigma$-range [5~\%, 20~\%]) at $0.5''$ and 
$\sim 45~\%$ ([30~\%, 60~\%]) at $1.5''$. From their data,
it is not clear whether or not they see in polarized flux a maximum 
at a separation of around $0.6''$ and a flux decrease inside. \cite{Thalmann2013} 
also fit the observations with model 
calculations but they adopt a radius of $r_0=40$~AU for the 
planetesimal ring and do not investigate models with larger $r_0$. 

The new SPHERE - ZIMPOL observations presented in this work provide a
very much improved polarimetric sensitivity which clearly reveals
a maximum in the polarization profile $P(x)$ 
at a projected separation of 0.60$\pm 0.06''$ (74$\pm 7$~AU).  
This maximum location is not compatible with the small ring radius of 
$r_0=40$~AU adopted by \cite{Thalmann2013}, probably, because they only 
fit the intensity profile which shows no features that could 
constrain the ring radius.

Our best-fit model is in good agreement with results of recent 
observations of HIP 79977 with ALMA. \cite{Lieman-Sifry2016} used 
the 1240 $ \mu$m continuum visibilities and derived basic geometrical 
parameters of the disk. They modelled the surface density of the disk with a
single power law $ r^{-1}$ extending from an inner to an outer radius 
and they derived $R_{\rm {inner}}=60^{+11}_{-13}$ AU without detecting 
an outer cut-off radius. This result confirms the large ring radius
$r_0$ found by us from the polarimetric profile. \cite{Lieman-Sifry2016} 
also measured a disk inclination $i > 84^\circ $ and $ PA = 115^{+3}_{-3} $ 
which are consistent with our and previous results.

\subsection{Diagnostic potential of polarimetry}

The scattered flux has been measured for more than 20 debris disks, mainly with the 
Hubble Space Telescope (HST)
\citep[e.g.,][]{Schneider2014, Schneider2016}. With ground-based observations the flux measurement for the 
scattered light from debris disks is very difficult 
because of the speckle noise introduced by the atmospheric 
turbulence. The polarized flux of the disk
$(F_{\rm pol})_{\rm disk}$ is much easier to determine, 
because it is a differential quantity which can be
distinguished from the unpolarized light from the bright
central star ($(F_{\rm pol})_{\ast}\approx 0$), even in the
presence of strong atmospheric speckles. 

For HIP 79977 the disk profile in polarized flux reveals a clear maximum which traces the
radial location of the disk ring. This information is
difficult to obtain from intensity imaging of edge-on disks, 
because the Stokes I disk profile is
dominated by the forward scattering dust in front of the
star and therefore the projected disk extension may not be visible.

The disk flux $F_{\rm disk}$ and the polarized flux $(F_{\rm pol})_{\rm disk}$
contain complementary information about the scattering dust.
The scattering angle dependence is strongly different
because forward and backward scattering produces no or only very 
little polarized flux. This applies also to the diffraction peak 
(or forward scattering peak) from large particles
$a>\lambda$, which is not or only slightly polarized. This means that
the polarized flux originates predominantly from scatterings with
scattering angles in the range $45^\circ-135^\circ$ and the polarized flux 
produced per scattering event can 
be approximated by an averaged particle parameter for the induced 
scattering polarization $p_m$ for the scattering angle 
of $90^\circ$ (see Sect.~\ref{phase f}).

New constraints on dust properties may be obtained if quantitative
polarimetric data of many debris disks can be collected. 
The dust grain size distribution and therefore the polarimetric
properties are expected to depend on the spectral type of 
the central star and different system ages may reveal 
evolutionary processes in the polarimetric properties of the 
scattering dust. Polarimetric parameters which can be 
quantified for the dust are scattering cross-section $\sigma_{\rm sca}$
or albedo, and parameters of the polarimetric scattering
function $p_m$ and $g_{\rm sca}$. A better understanding of 
the dust in debris disks would be very useful for interpretations
regarding the nature of their parent bodies which produced the observed
dust in a collisional cascade.
 
\begin{figure}  
   \centering
   \includegraphics[width=9cm]{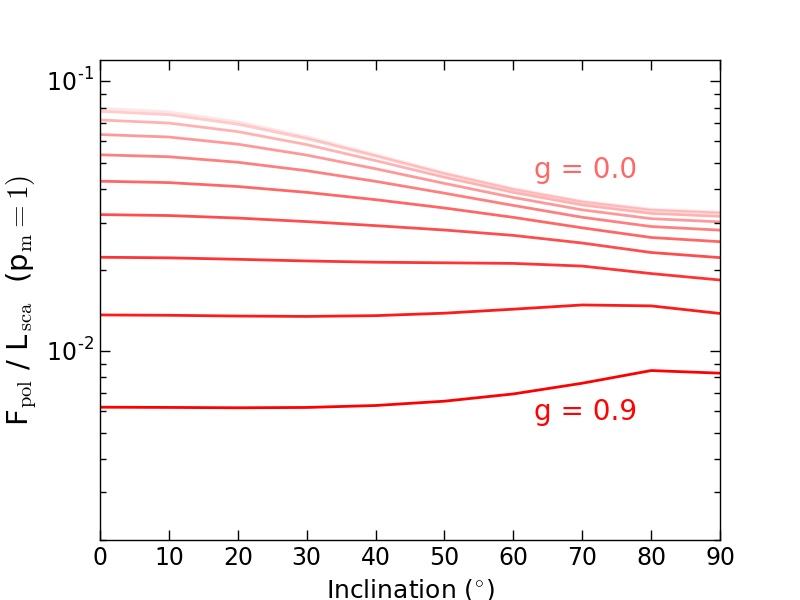}
               \caption{Ratio of polarized flux to the scattered light 
   luminosity for optically thin debris disks as a function of disk inclination 
  and scattering asymmetry parameter $g_{\rm sca}$ (plotted for $g_{\rm sca} = 0.0, 0.2, 0.3, ...,  0.9$). The ratio is independent of the disk geometry and follows from the
scattering phase functions as shown in Fig. \ref{Phase_functions}. \label{ratio pol}}
 \end{figure}
 
\subsubsection{Polarized flux and infrared excess}  

The reflectivity or scattering albedo of the dust in the debris disk of HIP 79977 can
be characterized by a comparison of the scattered polarized flux with
the IR excess luminosity which is a good measure for the dust absorption. 

In Sect. \ref{Contrast}, we derived the fractional polarized flux or
ratio of total polarized disk flux to the stellar flux 
$(F_{\rm pol})_{\rm disk}/F_{\rm \ast} = (5.5 \pm 0.9)\cdot 10^{-4} $ . This ratio
was obtained for the wide VBB filter near the peak
of the stellar energy distribution. Therefore, we can consider
the ratio $(F_{\rm pol})_{\rm disk}/F_{\rm \ast}$ as a good order of magnitude
estimate for the fractional polarized light luminosity of the disk expressed as 
$(L_{\mathrm{pol}})_{\rm disk}/L_{\ast}$. This statement 
considers also the fact that $(F_{\rm pol}(i))_{\rm disk}/(L_{\rm sca})_{\rm disk}$ depends very little,
less than a factor of two, upon the disk inclination.

The fractional infrared excess of HIP 79977 $L_{\mathrm{IR}}/L_{\ast}=5.21\cdot10^{-3}$ 
is given in \cite{Jang-Condell2015}. 
This yields the double ratio 
\begin{displaymath}
\Lambda = {(F_{\rm pol})_{\rm disk}/F_{\rm \ast} \over L_{\mathrm{IR}}/L_{\ast}} = 0.11 \pm 0.02,
\end{displaymath}  
where the uncertainty only accounts for uncertainty in the 
$(F_{\rm pol})_{\rm disk}/F_{\rm \ast}$-ratio derived in this paper.
The double ratio $\Lambda$ could be a good proxy for the 
ratio between the polarized luminosity and the IR-excess luminosity
of the disk
\begin{displaymath}
\Lambda={(F_{\mathrm{pol}})_{\rm disk}/F_{\ast}\over L_{\mathrm{IR}}/L_{\ast}} \approx {(L_{\mathrm{pol}})_{\rm disk}\over L_{\mathrm{IR}}} \,,
\end{displaymath}
if the wavelength and inclination dependence of the 
dust scattering can be neglected. 
It is emphasized, that neglecting the wavelength dependence of the 
polarized flux of the disk may not be an acceptable simplification 
for some cases, for example, for near-IR polarimetry of disks around 
A-stars, which emit most of their radiation in the UV-visual 
spectral region. 

Therefore, $\Lambda$ is an observational parameter that 
depends, like the scattering albedo, on the ratio 
between dust scattering cross-section
$\sigma_{\rm sca}$ and absorption $\kappa$, and parameters of the polarimetric phase function $p_m$ and $g_{\rm sca}$ as
\begin{equation}
\Lambda \propto {\sigma_{\rm sca}(\lambda)\over \kappa} \cdot f(p_m(\lambda),g_{\rm sca}(\lambda), i)\,.
\end{equation}

The inclination dependence of $(F_{\rm pol})_{\rm disk}$ is illustrated in 
Fig.~\ref{ratio pol}, which shows the polarized flux $(F_{\rm pol})_{\rm disk}$ (expressed
per steradian) with respect to the scattered light 
luminosity $L_{\rm sca}$ excluding the
diffracted light. The scattered light interacts with the surface of the
dust particles, and, therefore, the asymmetry parameter $g_{\rm sca}$, which we
introduced for the polarized light, is also adopted for the intensity of the
scattered light as a first approximation (see Sect.~\ref{phase f}). 

For isotropic scattering $g=0$ and maximum polarization ($p_m=1$), the 
ratio of polarized flux to scattering luminosity is 
$(F_{\rm pol})_{\rm disk}/L_{\rm sca}=1/4\pi$ for $i=0^\circ$ 
because the scattering angle for a pole-on disk is $90^\circ$ throughout and the
radiation is 100~\% polarized. For larger inclinations (for $g_{\rm sca}=0$)
the ratio is smaller, because there is more forward and backward scattering
which produces less polarization. The ratio between a pole-on and edge-on disk is
$F_{\rm pol}(i=90^\circ)_{\rm disk}/F_{\rm pol}(i=0^\circ)= 0.41 $. We note that pure
Rayleigh scattering is different, because it is not isotropic.

For strong forward scattering $g_{\rm sca}\rightarrow 1$ the amount of 
polarized light is reduced with respect to the scattered intensity 
(see Fig.~\ref{Phase_functions}) or the disk luminosity in scattered light 
(without diffraction). For $g\approx 0.6-0.8$ the polarized 
flux $(F_{\rm pol})_{\rm disk}$ is therefore almost independent of the 
disk inclination, and for $g\approx 0.8$, edge-on disks are even brighter in $(F_{\rm pol})_{\rm disk}$ 
than pole-on disks because so much more light is scattered in 
forward directions. 

Figure~\ref{ratio pol} is independent of the radial mass distribution
for rotationally symmetric, flat, optically thin disks 
and a given scattering phase matrix. The inclination can usually be 
determined easily. More difficult is the 
determination of
the scattering asymmetry parameter $g_{\rm sca}$, at least for edge-on disks and
pole-on disks. For HIP 79977, the 1-$\sigma$ uncertainty range for 
$g_{\rm sca}$ is [0.2,0.7], and this leaves 
an uncertainty of about a factor 1.5 (see Fig.~\ref{ratio pol}) for the 
$(F_{\rm pol})_{\rm disk}/L_{\rm sca}$ ratio
determination. In addition, there is also the factor $p_m$, which needs 
to be known to constrain the mean scattering albedo of the dust in 
debris disks.     

Clearly, there is not a straight-forward way to derive a value for 
$f(p_m(\lambda),g_{\rm sca}(\lambda), i)$ from a polarimetric observation 
of a single disk. However, we can expect progress if the polarized flux 
is derived for several disks with different inclinations, including
cases where the $g_{\rm sca}$-asymmetry parameter can be well defined.

Also of great value would be polarimetric observation, for which well
calibrated HST intensity images are available to complement 
the $F_{\rm disk}/F_{\ast} - L_{\mathrm{IR}}/L_{\ast}$-plot 
of \citet[][their Fig.~8]{Schneider2014} with an equivalent plot for 
the polarized disk flux $(F_{\mathrm{pol}})_{\rm disk}$ and constrain differences 
between scattered intensity and polarized intensity of debris disks.
 
Up to now there exists only few polarized flux $(F_{\rm pol})_{\rm disk}$
measurements for debris disks and therefore it is difficult
to compare different disks. More data will
become available soon from the new polarimetric high-contrast observing modes of 
SPHERE and other similar instruments (e.g., GPI, HiCIAO). 

 \begin{figure}  
   \centering
   \includegraphics[width=9cm]{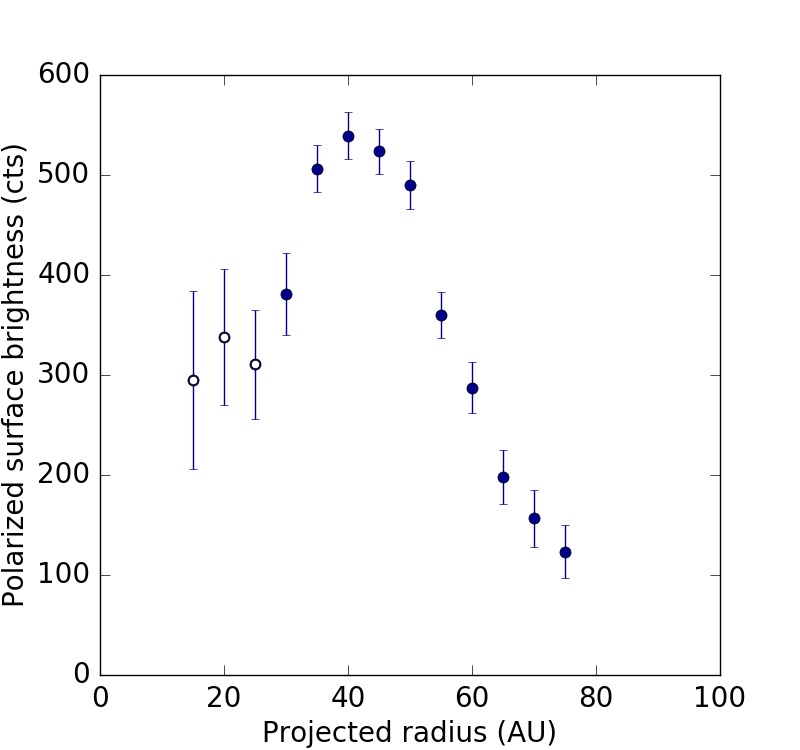}
               \caption{Polarized flux profile $\langle P|x|\rangle$ of the AU Mic debris 
disk derived from the polarimetric profiles of \citet[][their Fig.~4]{Graham2007}. 
The bins are 5 AU wide and error bars are obtained from error propagation including systematic uncertainties. 
Open circles are derived from noisy data without considering the bias effect for the determination of the 
fractional polarization \citep{Clarke1983}. \label{AUMic}}
 \end{figure}
 
\subsubsection{Comparison with the edge-on disk AU Mic}
The disk around the nearby (9.9 pc) low-mass star AU Mic (M1Ve)
is a very good example of a previous high-quality study
of the polarization of an edge-on debris disk. 
HST imaging for this target is presented by \cite{Krist2005} and \cite{Schneider2014} and imaging polarimetry is
described in \cite{Graham2007}. 

\citet{Krist2005} measure from their AU Mic intensity image a disk width
which increases with apparent separations qualitatively similar to the
behavior measured from our polarized intensity image for HIP 79977
(Fig.~\ref{SBprofile}). \citet{Graham2007} present a profile
of the fractional polarization $p(x)$ along the disk spine together
with an intensity profile $I(x)$ for the F606W filter-band ($\lambda_{\rm c}$ = 0.590 $\mu$m,  $\Delta\lambda= 0.230$ $\mu$m). We obtained their profiles
(J. Graham, personal communication) and constructed for AU Mic a mean
disk profile $\langle P|x|\rangle = (p(x<0)I(x<0)+p(x>0)I(x>0))/2$
given in Fig.~\ref{AUMic}. This profile shows
a maximum value and essentially no polarized flux close to the star, again very similar 
to HIP 79977 (Fig.~\ref{SBprofile}). 

In AU Mic, the peak of the polarized flux is at 40 AU 
and this coincides well with the outer edge of the dust belt
seen in the ALMA 1.3 mm dust continuum, which probably traces the outer
edge of the suspected ``birth ring'' of colliding planetesimals \citep{MacGregor2013}.
This finding for AU Mic supports our interpretation of HIP 79977 data, that 
the measured maximum polarization at $r\sim$75~AU represents well the ``birth ring'' radius. 

The HST polarimetry of \citet{Graham2007} is not flux calibrated and therefore
we used the F606W imaging of AU Mic of \citet{Krist2005}. They provide a
calibrated SB profile of the disk spine and the disk widths at 
different separations from which we derive a calibrated intensity
profile $I(x)$. With this, we calibrate the polarimetry of \citet{Graham2007}
and derive for the AU Mic disc in the F606W filter a total polarized flux 
of $0.31 \pm 0.11$ mJy. This includes the disk regions from
$1''$ to 11$''$ on both sides but not the innermost arcsec. The calculated
polarized flux relative to the stellar flux is 
$(F_{\rm pol})_{\rm disk}/F_{\rm \ast} \approx (2.41 \pm 0.84)\cdot 10^{-4}$. 

With the fractional infrared luminosity $L_{\mathrm{IR}}/L_{\ast}=4.4\cdot10^{-4}$ \citep{Plavchan2009}
we obtain a $\Lambda$-parameter equal to $0.55 \pm 0.19$ for AU Mic. It is interesting to
note that the debris dust in AU Mic produces approximately five times as much scattering 
polarization when compared to HIP 79977, if we compare the fractional polarized flux in the F606W filter 
to the fractional infrared excess. 

The interpretation of this difference is not clear. One possibility is, that the blue color of the disk around AU Mic is caused by a surplus of very small grains when compared to other debris disks \citep{Krist2005}. Roughly, the maximum blow-out size scales like the ratio $L_\ast / M_\ast$ between the stellar luminosity and the stellar mass \citep{Burns1979}. For AU Mic this ratio is about an order of magnitude smaller than for HIP 79977. When disregarding possible small differences in grain properties it implies that the minimum grain size in the disk around AU Mic is about $a_{\rm min} \approx 0.1\, \mu$m, in rough agreement with results obtained by \cite{Schueppler2015}. 
The corresponding value is $a_{\rm min} \approx 0.9\, \mu$m for HIP 79977. For this reason, the polarized flux derived for the HST F606W might not represent a good wavelength average for the scattering polarization in AU Mic and the derived $\Lambda$-parameter must therefore be interpreted with caution.

\subsubsection{Dust scattering properties and disk model} 

In this work the observed polarized flux from the debris disk 
in HIP 79977 is fitted with disk models. For this, the parameters 
describing the dust scattering were restricted to 
$g_{\rm sca}$ for the scattering phase angle dependence for polarized light 
and $p_m$ for the amount of polarized light produced by the scattering.    
Unfortunately, it was not possible to constrain these 
parameters well with the modeling. 

The maximum fractional polarization produced per scattering 
$p_m$ can only be constrained if the intensity $F_{\mathrm{disk}}$ of the 
scattered light from the disk can also be accurately measured. The intensity
signal of HIP 79977 is clearly detected with SPHERE-ZIMPOL but this signal is 
strongly affected by self-subtraction effects 
of the ADI procedure which are hard to quantify. Therefore, a determination
of $F_{\mathrm{disk}}$ and $p_m$ needs a more accurate stellar PSF subtraction technique, 
as is possible with HST. 

The asymmetry parameter for the polarized scattered flux $g_{\rm sca}$ is
also not well constrained because of the edge-on configuration of
the HIP 79977 disk. We clearly see the expected polarization minimum for 
forward and backward scattering at small angular separation from 
the star. However, the more subtle asymmetry parameter $g_{\rm sca}$ is not well 
defined. From the modeling of the polarization profile along the disk spine $\langle P|x|\rangle$,
it is not possible to disentangle the parameters  
$r_0,\alpha_{\rm in},\alpha_{\rm out}$ for the radial 
distribution of the dust from the scattering asymmetry
parameter $g_{\rm sca}$. Our data show at least that $g_{\rm sca}>0.2$, that is, 
much more polarized flux is produced in the forward scattering direction because the disk spine in
polarized flux is on the same side of the central star, like the
intensity spine caused by forward diffraction.

A more accurate determination of $g_{\rm sca}$ will be possible for
debris disks with a slightly smaller inclination, where the azimuthal 
dependence of the polarized flux can be defined.

\section{Summary}
In this paper, we present SPHERE-ZIMPOL images of polarized light 
of the debris disk around HIP 79977 
in the 590-880 nm wavelength range using differential polarimetry and an intensity
image extracted with angular differential imaging using the LOCI-algorithm. 
We have characterized and analyzed the disk structure, mainly
based on the polarized flux image and obtained the following results:
\begin{itemize}
\item  The images show a nearly edge-on disk 
extending from less than $0.1''$ out to the edge of the detector 
at $1.8'' (\sim$225 AU): they unveil regions close to the star which were
hidden by residual atmospheric speckles in previous data \citep[cf.][]{Thalmann2013}.
\item For the PA of the disk, we measure 
$\theta_{\rm disk} = 114.5^{\circ} \pm 0.6^{\circ}$, which is in a good 
agreement with the value reported by \cite{Thalmann2013} and \cite{Lieman-Sifry2016}.
\item From our polarimetric data, we derive disk cross-sections 
perpendicular to the disk midplane. The peaks of the perpendicular
profiles are slightly offset ($\approx 30-60$~mas) because we are seeing
a strong flux asymmetry between the front and back sides of a highly
inclined disk. At small apparent separations ($r < 0.5''$) the
profiles have a super-exponential drop-off pointing to a well defined concentration of large dust particles
in the midplane of the inner disk. The disk width (FWHM) is increasing 
systematically from ${\rm FWHM}\approx 0.1''$ (12 AU) to $0.3'' - 0.5''$, 
when going from a separation of $0.2''$ to $>1''$. The growth of the 
profile scale height could be caused by a radial blow-out of small grains. 
\item The disk surface brightness profile in polarized light along 
the disk spine is symmetric on the ESE and WNW sides. There is a clear maximum 
of ${\rm SB}_{\rm max}=16.2\,{\rm mag}\,{\rm arcsec}^{-2}$ 
between $0.2''$ and $0.5''$ where the surface brightness 
contrast with respect to the central star is $7.64\,{\rm mag}\,{\rm arcsec}^{-2}$. There is a clear 
minimum of SB closer to the star ($r<0.2''$), because no or only little 
polarization is produced by forward scattering in 
the disk section lying in front of the star which dominates the signal 
in the intensity image. The geometric structure of the disk seen in polarized light is
consistent with intensity images taken with SPHERE-ZIMPOL 
and literature data. Unfortunately, it is difficult and we were not able to derive
a high-quality intensity profile for the disk and therefore we
could not make a quantitative comparison between intensity and polarized flux.
\item The disk profile in polarized flux for HIP 79977 shows a clear maximum at
the projected separation of $74\pm 7$~AU. This seems to be a good
measure of the belt radius for this edge-on debris disk. 
\item The disk total polarized flux amounts to 
$mp_{\mathrm{disk}}(\mathrm{VBB})$ = 16.6 $\pm$ 0.3 which is 
$(F_{\rm pol})_{\rm disk}/F_{\rm \ast} = (5.5 \pm 0.9)\cdot 10^{-4} $. 
\item The ratio $\Lambda \approx 0.11$ compares the disk polarized flux 
with the disk infrared excess. We emphasize the value of this parameter 
for the characterization of the scattering albedo of the dust particles. 
For comparative purposes, we derive the ratio $\Lambda \approx 0.55$ for 
the edge-on debris disk around the M star AU Mic based on the previous HST observations of this target.
\end{itemize}

The dust distribution of the disk around HIP 79977 was modelled with 
a 3D rotationally symmetric belt of radius $r_0$, with radial 
power laws for the density fall-off inside and outside $r_0$ and 
an exponential function in vertical density distribution. 
A large grid of models was fitted to the data and we derive
the best disk parameters using a $\chi^2$ optimization technique.
This model analysis yields a disk with an inclination of $i\approx 85^\circ$ and 
a belt radius $r_0 \approx 73$~AU and a grain density distribution with a steep power law index 
$\alpha=5$ inside $r_0$ and a more shallow index $\alpha=-2.5$ 
outside $r_0$. The derived scattering asymmetry parameter lies between 
$g_{\rm sca}=0.2$ and 0.6 (forward scattering) using an adopted angle-dependence for the fractional polarization 
like for Rayleigh scattering.

 \begin{acknowledgements}
We thank James Graham for providing the AU Mic disk profiles derived from
HST data and published in Graham et al. (2007, Fig.~4) in electronic form. 
We would also like to thank the referee for many thoughtful comments 
which helped to improve this paper. 
This work is supported by the Swiss National Science Foundation
through grant number 200020 - 162630. 
SPHERE is an instrument designed and built by a consortium consisting
of IPAG (Grenoble, France), MPIA (Heidelberg, Germany), LAM (Marseille,
France), LESIA (Paris, France), Laboratoire Lagrange (Nice, France), 
INAF – Osservatorio di Padova (Italy), Observatoire de Gen`eve 
(Switzerland), ETH Zurich (Switzerland), NOVA (Netherlands), ONERA (France) 
and ASTRON (Netherlands), in collaboration with ESO. 
SPHERE was funded by ESO, with additional contributions from CNRS (France), 
MPIA (Germany), INAF (Italy), FINES (Switzerland) and NOVA (Netherlands). 
SPHERE also received funding from the European Commission Sixth and 
Seventh Framework Programmes as part of the Optical Infrared 
Coordination Network for Astronomy (OPTICON)
under grant number RII3-Ct-2004-001566 for FP6 (2004–2008), grant number
226604 for FP7 (2009–2012) and grant number 312430 for FP7 (2013–2016). J.O acknowledges support from ALMA/Conicyt Project 31130027, and from the Millennium Nucleus RC130007 (Chilean Ministry of Economy).
      \end{acknowledgements}

\bibliographystyle{aa} 

\bibliography{reference.bib} 

\begin{thebibliography}{74}
\expandafter\ifx\csname natexlab\endcsname\relax\def\natexlab#1{#1}\fi

\bibitem[{{Ahmic} {et~al.}(2009){Ahmic}, {Croll}, \& {Artymowicz}}]{Ahmic2009}
{Ahmic}, M., {Croll}, B., \& {Artymowicz}, P. 2009, \apj, 705, 529

\bibitem[{{Artymowicz} {et~al.}(1989){Artymowicz}, {Burrows}, \&
  {Paresce}}]{Artymowicz1989}
{Artymowicz}, P., {Burrows}, C., \& {Paresce}, F. 1989, \apj, 337, 494

\bibitem[{{Augereau} {et~al.}(2001){Augereau}, {Nelson}, {Lagrange},
  {Papaloizou}, \& {Mouillet}}]{Augereau2001}
{Augereau}, J.~C., {Nelson}, R.~P., {Lagrange}, A.~M., {Papaloizou}, J.~C.~B.,
  \& {Mouillet}, D. 2001, \aap, 370, 447

\bibitem[{{Aumann} {et~al.}(1984){Aumann}, {Beichman}, {Gillett}, {de Jong},
  {Houck}, {Low}, {Neugebauer}, {Walker}, \& {Wesselius}}]{Aumann1984}
{Aumann}, H.~H., {Beichman}, C.~A., {Gillett}, F.~C., {et~al.} 1984, \apjl,
  278, L23

\bibitem[{{Backman} \& {Paresce}(1993)}]{Backman1993}
{Backman}, D.~E. \& {Paresce}, F. 1993, in Protostars and Planets III, ed.
  E.~H. {Levy} \& J.~I. {Lunine}, 1253--1304

\bibitem[{{Bazzon} {et~al.}(2012){Bazzon}, {Gisler}, {Roelfsema}, {Schmid},
  {Pragt}, {Elswijk}, {de Haan}, {Downing}, {Salasnich}, {Pavlov}, {Beuzit},
  {Dohlen}, {Mouillet}, \& {Wildi}}]{Bazzon2012}
{Bazzon}, A., {Gisler}, D., {Roelfsema}, R., {et~al.} 2012, in \procspie, Vol.
  8446, Ground-based and Airborne Instrumentation for Astronomy IV, 844693

\bibitem[{{Beuzit} {et~al.}(2008){Beuzit}, {Feldt}, {Dohlen}, {Mouillet},
  {Puget}, {Wildi}, {Abe}, {Antichi}, {Baruffolo}, {Baudoz}, {Boccaletti},
  {Carbillet}, {Charton}, {Claudi}, {Downing}, {Fabron}, {Feautrier},
  {Fedrigo}, {Fusco}, {Gach}, {Gratton}, {Henning}, {Hubin}, {Joos}, {Kasper},
  {Langlois}, {Lenzen}, {Moutou}, {Pavlov}, {Petit}, {Pragt}, {Rabou}, {Rigal},
  {Roelfsema}, {Rousset}, {Saisse}, {Schmid}, {Stadler}, {Thalmann}, {Turatto},
  {Udry}, {Vakili}, \& {Waters}}]{Beuzit2008}
{Beuzit}, J.-L., {Feldt}, M., {Dohlen}, K., {et~al.} 2008, in \procspie, Vol.
  7014, Ground-based and Airborne Instrumentation for Astronomy II, 701418

\bibitem[{{Buenzli} {et~al.}(2010){Buenzli}, {Thalmann}, {Vigan}, {Boccaletti},
  {Chauvin}, {Augereau}, {Meyer}, {M{\'e}nard}, {Desidera}, {Messina},
  {Henning}, {Carson}, {Montagnier}, {Beuzit}, {Bonavita}, {Eggenberger},
  {Lagrange}, {Mesa}, {Mouillet}, \& {Quanz}}]{Buenzli2010}
{Buenzli}, E., {Thalmann}, C., {Vigan}, A., {et~al.} 2010, \aap, 524, L1

\bibitem[{{Burns} {et~al.}(1979){Burns}, {Lamy}, \& {Soter}}]{Burns1979}
{Burns}, J.~A., {Lamy}, P.~L., \& {Soter}, S. 1979, \icarus, 40, 1

\bibitem[{{Canovas} {et~al.}(2015){Canovas}, {M{\'e}nard}, {de Boer}, {Pinte},
  {Avenhaus}, \& {Schreiber}}]{Canovas2015}
{Canovas}, H., {M{\'e}nard}, F., {de Boer}, J., {et~al.} 2015, \aap, 582, L7

\bibitem[{{Chen} {et~al.}(2011){Chen}, {Mamajek}, {Bitner}, {Pecaut}, {Su}, \&
  {Weinberger}}]{Chen2011}
{Chen}, C.~H., {Mamajek}, E.~E., {Bitner}, M.~A., {et~al.} 2011, \apj, 738, 122

\bibitem[{{Chen} {et~al.}(2006){Chen}, {Sargent}, {Bohac}, {Kim},
  {Leibensperger}, {Jura}, {Najita}, {Forrest}, {Watson}, {Sloan}, \&
  {Keller}}]{Chen2006}
{Chen}, C.~H., {Sargent}, B.~A., {Bohac}, C., {et~al.} 2006, \apjs, 166, 351

\bibitem[{{Clarke} {et~al.}(1983){Clarke}, {Stewart}, {Schwarz}, \&
  {Brooks}}]{Clarke1983}
{Clarke}, D., {Stewart}, B.~G., {Schwarz}, H.~E., \& {Brooks}, A. 1983, \aap,
  126, 260

\bibitem[{{Cutri} {et~al.}(2003){Cutri}, {Skrutskie}, {van Dyk}, {Beichman},
  {Carpenter}, {Chester}, {Cambresy}, {Evans}, {Fowler}, {Gizis}, {Howard},
  {Huchra}, {Jarrett}, {Kopan}, {Kirkpatrick}, {Light}, {Marsh}, {McCallon},
  {Schneider}, {Stiening}, {Sykes}, {Weinberg}, {Wheaton}, {Wheelock}, \&
  {Zacarias}}]{Cutri2003}
{Cutri}, R.~M., {Skrutskie}, M.~F., {van Dyk}, S., {et~al.} 2003, VizieR Online
  Data Catalog, 2246

\bibitem[{{Debes} {et~al.}(2013){Debes}, {Jang-Condell}, {Weinberger},
  {Roberge}, \& {Schneider}}]{Debes2013}
{Debes}, J.~H., {Jang-Condell}, H., {Weinberger}, A.~J., {Roberge}, A., \&
  {Schneider}, G. 2013, \apj, 771, 45

\bibitem[{{Debes} {et~al.}(2008){Debes}, {Weinberger}, \&
  {Schneider}}]{Debes2008}
{Debes}, J.~H., {Weinberger}, A.~J., \& {Schneider}, G. 2008, \apjl, 673, L191

\bibitem[{{Dohlen} {et~al.}(2006){Dohlen}, {Beuzit}, {Feldt}, {Mouillet},
  {Puget}, {Antichi}, {Baruffolo}, {Baudoz}, {Berton}, {Boccaletti},
  {Carbillet}, {Charton}, {Claudi}, {Downing}, {Fabron}, {Feautrier},
  {Fedrigo}, {Fusco}, {Gach}, {Gratton}, {Hubin}, {Kasper}, {Langlois},
  {Longmore}, {Moutou}, {Petit}, {Pragt}, {Rabou}, {Rousset}, {Saisse},
  {Schmid}, {Stadler}, {Stamm}, {Turatto}, {Waters}, \& {Wildi}}]{Dohlen2006}
{Dohlen}, K., {Beuzit}, J.-L., {Feldt}, M., {et~al.} 2006, in \procspie, Vol.
  6269, Society of Photo-Optical Instrumentation Engineers (SPIE) Conference
  Series, 62690Q

\bibitem[{{Draper} {et~al.}(2016){Draper}, {Duch{\^e}ne}, {Millar-Blanchaer},
  {Matthews}, {Wang}, {Kalas}, {Graham}, {Padgett}, {Ammons}, {Bulger}, {Chen},
  {Chilcote}, {Doyon}, {Fitzgerald}, {Follette}, {Gerard}, {Greenbaum},
  {Hibon}, {Hinkley}, {Macintosh}, {Ingraham}, {Lafreni{\`e}re}, {Marchis},
  {Marois}, {Nielsen}, {Oppenheimer}, {Patel}, {Patience}, {Perrin}, {Pueyo},
  {Rajan}, {Rameau}, {Sivaramakrishnan}, {Vega}, {Ward-Duong}, \&
  {Wolff}}]{Draper2016}
{Draper}, Z.~H., {Duch{\^e}ne}, G., {Millar-Blanchaer}, M.~A., {et~al.} 2016,
  \apj, 826, 147

\bibitem[{{Duch{\^e}ne} {et~al.}(2014){Duch{\^e}ne}, {Arriaga}, {Wyatt},
  {Kennedy}, {Sibthorpe}, {Lisse}, {Holland}, {Wisniewski}, {Clampin}, {Kalas},
  {Pinte}, {Wilner}, {Booth}, {Horner}, {Matthews}, \& {Greaves}}]{Duchene2014}
{Duch{\^e}ne}, G., {Arriaga}, P., {Wyatt}, M., {et~al.} 2014, \apj, 784, 148

\bibitem[{{ESA}(1997)}]{ESA1997}
{ESA}, ed. 1997, ESA Special Publication, Vol. 1200, {The HIPPARCOS and TYCHO
  catalogues. Astrometric and photometric star catalogues derived from the ESA
  HIPPARCOS Space Astrometry Mission}

\bibitem[{{Fusco} {et~al.}(2014){Fusco}, {Sauvage}, {Petit}, {Costille},
  {Dohlen}, {Mouillet}, {Beuzit}, {Kasper}, {Suarez}, {Soenke}, {Fedrigo},
  {Downing}, {Baudoz}, {Sevin}, {Perret}, {Barrufolo}, {Salasnich}, {Puget},
  {Feautrier}, {Rochat}, {Moulin}, {Deboulb{\'e}}, {Hugot}, {Vigan}, {Mawet},
  {Girard}, \& {Hubin}}]{Fusco2014}
{Fusco}, T., {Sauvage}, J.-F., {Petit}, C., {et~al.} 2014, in \procspie, Vol.
  9148, Adaptive Optics Systems IV, 91481U

\bibitem[{{Gaia Collaboration}(2016)}]{GaiaCollaboration2016}
{Gaia Collaboration}. 2016, VizieR Online Data Catalog, 1337

\bibitem[{{Garufi} {et~al.}(2016){Garufi}, {Quanz}, {Schmid}, {Mulders},
  {Avenhaus}, {Boccaletti}, {Ginski}, {Langlois}, {Stolker}, {Augereau},
  {Benisty}, {Lopez}, {Dominik}, {Gratton}, {Henning}, {Janson}, {M{\'e}nard},
  {Meyer}, {Pinte}, {Sissa}, {Vigan}, {Zurlo}, {Bazzon}, {Buenzli}, {Bonnefoy},
  {Brandner}, {Chauvin}, {Cheetham}, {Cudel}, {Desidera}, {Feldt}, {Galicher},
  {Kasper}, {Lagrange}, {Lannier}, {Maire}, {Mesa}, {Mouillet}, {Peretti},
  {Perrot}, {Salter}, \& {Wildi}}]{Garufi2016}
{Garufi}, A., {Quanz}, S.~P., {Schmid}, H.~M., {et~al.} 2016, \aap, 588, A8

\bibitem[{{Gledhill} {et~al.}(1991){Gledhill}, {Scarrott}, \&
  {Wolstencroft}}]{Gledhill1991}
{Gledhill}, T.~M., {Scarrott}, S.~M., \& {Wolstencroft}, R.~D. 1991, \mnras,
  252, 50P

\bibitem[{{Golimowski} {et~al.}(2006){Golimowski}, {Ardila}, {Krist},
  {Clampin}, {Ford}, {Illingworth}, {Bartko}, {Ben{\'{\i}}tez}, {Blakeslee},
  {Bouwens}, {Bradley}, {Broadhurst}, {Brown}, {Burrows}, {Cheng}, {Cross},
  {Demarco}, {Feldman}, {Franx}, {Goto}, {Gronwall}, {Hartig}, {Holden},
  {Homeier}, {Infante}, {Jee}, {Kimble}, {Lesser}, {Martel}, {Mei},
  {Menanteau}, {Meurer}, {Miley}, {Motta}, {Postman}, {Rosati}, {Sirianni},
  {Sparks}, {Tran}, {Tsvetanov}, {White}, {Zheng}, \& {Zirm}}]{Golimowski2006}
{Golimowski}, D.~A., {Ardila}, D.~R., {Krist}, J.~E., {et~al.} 2006, \aj, 131,
  3109

\bibitem[{{Graham} {et~al.}(2007){Graham}, {Kalas}, \& {Matthews}}]{Graham2007}
{Graham}, J.~R., {Kalas}, P.~G., \& {Matthews}, B.~C. 2007, \apj, 654, 595

\bibitem[{{Henyey} \& {Greenstein}(1941)}]{Henyey1941}
{Henyey}, L.~G. \& {Greenstein}, J.~L. 1941, \apj, 93, 70

\bibitem[{{Hinkley} {et~al.}(2009){Hinkley}, {Oppenheimer}, {Soummer},
  {Brenner}, {Graham}, {Perrin}, {Sivaramakrishnan}, {Lloyd}, {Roberts}, \&
  {Kuhn}}]{Hinkley2009}
{Hinkley}, S., {Oppenheimer}, B.~R., {Soummer}, R., {et~al.} 2009, \apj, 701,
  804

\bibitem[{{H{\o}g} {et~al.}(2000){H{\o}g}, {Fabricius}, {Makarov}, {Urban},
  {Corbin}, {Wycoff}, {Bastian}, {Schwekendiek}, \& {Wicenec}}]{Hog2000}
{H{\o}g}, E., {Fabricius}, C., {Makarov}, V.~V., {et~al.} 2000, \aap, 355, L27

\bibitem[{{Jang-Condell} {et~al.}(2015){Jang-Condell}, {Chen}, {Mittal},
  {Manoj}, {Watson}, {Lisse}, {Nesvold}, \& {Kuchner}}]{Jang-Condell2015}
{Jang-Condell}, H., {Chen}, C.~H., {Mittal}, T., {et~al.} 2015, \apj, 808, 167

\bibitem[{{Kalas} {et~al.}(2008){Kalas}, {Graham}, {Chiang}, {Fitzgerald},
  {Clampin}, {Kite}, {Stapelfeldt}, {Marois}, \& {Krist}}]{Kalas2008}
{Kalas}, P., {Graham}, J.~R., {Chiang}, E., {et~al.} 2008, Science, 322, 1345

\bibitem[{{Kalas} {et~al.}(2005){Kalas}, {Graham}, \& {Clampin}}]{Kalas2005}
{Kalas}, P., {Graham}, J.~R., \& {Clampin}, M. 2005, \nat, 435, 1067

\bibitem[{{Kasper} {et~al.}(2012){Kasper}, {Beuzit}, {Feldt}, {Dohlen},
  {Mouillet}, {Puget}, {Wildi}, {Abe}, {Baruffolo}, {Baudoz}, {Bazzon},
  {Boccaletti}, {Brast}, {Buey}, {Chesneau}, {Claudi}, {Costille},
  {Delboulb{\'e}}, {Desidera}, {Dominik}, {Dorn}, {Downing}, {Feautrier},
  {Fedrigo}, {Fusco}, {Girard}, {Giro}, {Gluck}, {Gonte}, {Gojak}, {Gratton},
  {Henning}, {Hubin}, {Lagrange}, {Langlois}, {Mignant}, {Lizon}, {Lilley},
  {Madec}, {Magnard}, {Martinez}, {Mawet}, {Mesa}, {M{\"u}ller-Nilsson},
  {Moulin}, {Moutou}, {O'Neal}, {Pavlov}, {Perret}, {Petit}, {Popovic},
  {Pragt}, {Rabou}, {Rochat}, {Roelfsema}, {Salasnich}, {Sauvage}, {Schmid},
  {Schuhler}, {Sevin}, {Siebenmorgen}, {Soenke}, {Stadler}, {Suarez},
  {Turatto}, {Udry}, {Vigan}, \& {Zins}}]{Kasper2012}
{Kasper}, M., {Beuzit}, J.-L., {Feldt}, M., {et~al.} 2012, The Messenger, 149,
  17

\bibitem[{{Krist} {et~al.}(2005){Krist}, {Ardila}, {Golimowski}, {Clampin},
  {Ford}, {Illingworth}, {Hartig}, {Bartko}, {Ben{\'{\i}}tez}, {Blakeslee},
  {Bouwens}, {Bradley}, {Broadhurst}, {Brown}, {Burrows}, {Cheng}, {Cross},
  {Demarco}, {Feldman}, {Franx}, {Goto}, {Gronwall}, {Holden}, {Homeier},
  {Infante}, {Kimble}, {Lesser}, {Martel}, {Mei}, {Menanteau}, {Meurer},
  {Miley}, {Motta}, {Postman}, {Rosati}, {Sirianni}, {Sparks}, {Tran},
  {Tsvetanov}, {White}, \& {Zheng}}]{Krist2005}
{Krist}, J.~E., {Ardila}, D.~R., {Golimowski}, D.~A., {et~al.} 2005, \aj, 129,
  1008

\bibitem[{{Lafreni{\`e}re} {et~al.}(2007){Lafreni{\`e}re}, {Marois}, {Doyon},
  {Nadeau}, \& {Artigau}}]{Lafreniere2007}
{Lafreni{\`e}re}, D., {Marois}, C., {Doyon}, R., {Nadeau}, D., \& {Artigau},
  {\'E}. 2007, \apj, 660, 770

\bibitem[{{Lagrange} {et~al.}(2010){Lagrange}, {Bonnefoy}, {Chauvin}, {Apai},
  {Ehrenreich}, {Boccaletti}, {Gratadour}, {Rouan}, {Mouillet}, {Lacour}, \&
  {Kasper}}]{Lagrange2010}
{Lagrange}, A.-M., {Bonnefoy}, M., {Chauvin}, G., {et~al.} 2010, Science, 329,
  57

\bibitem[{{Lieman-Sifry} {et~al.}(2016){Lieman-Sifry}, {Hughes}, {Carpenter},
  {Gorti}, {Hales}, \& {Flaherty}}]{Lieman-Sifry2016}
{Lieman-Sifry}, J., {Hughes}, A.~M., {Carpenter}, J.~M., {et~al.} 2016, \apj,
  828, 25

\bibitem[{{MacGregor} {et~al.}(2013){MacGregor}, {Wilner}, {Rosenfeld},
  {Andrews}, {Matthews}, {Hughes}, {Booth}, {Chiang}, {Graham}, {Kalas},
  {Kennedy}, \& {Sibthorpe}}]{MacGregor2013}
{MacGregor}, M.~A., {Wilner}, D.~J., {Rosenfeld}, K.~A., {et~al.} 2013, \apjl,
  762, L21

\bibitem[{{Marois} {et~al.}(2006){Marois}, {Lafreni{\`e}re}, {Doyon},
  {Macintosh}, \& {Nadeau}}]{Marois2006}
{Marois}, C., {Lafreni{\`e}re}, D., {Doyon}, R., {Macintosh}, B., \& {Nadeau},
  D. 2006, \apj, 641, 556

\bibitem[{{Marois} {et~al.}(2008){Marois}, {Macintosh}, {Barman}, {Zuckerman},
  {Song}, {Patience}, {Lafreni{\`e}re}, \& {Doyon}}]{Marois2008}
{Marois}, C., {Macintosh}, B., {Barman}, T., {et~al.} 2008, Science, 322, 1348

\bibitem[{{Milli} {et~al.}(2017){Milli}, {Vigan}, {Mouillet}, {Lagrange},
  {Augereau}, {Pinte}, {Mawet}, {Schmid}, {Boccaletti}, {Matr{\`a}}, {Kral},
  {Ertel}, {Chauvin}, {Bazzon}, {M{\'e}nard}, {Beuzit}, {Thalmann}, {Dominik},
  {Feldt}, {Henning}, {Min}, {Girard}, {Galicher}, {Bonnefoy}, {Fusco}, {de
  Boer}, {Janson}, {Maire}, {Mesa}, {Schlieder}, \& {SPHERE
  Consortium}}]{Milli2017}
{Milli}, J., {Vigan}, A., {Mouillet}, D., {et~al.} 2017, \aap, 599, A108

\bibitem[{{Min} {et~al.}(2010){Min}, {Kama}, {Dominik}, \& {Waters}}]{Min2010}
{Min}, M., {Kama}, M., {Dominik}, C., \& {Waters}, L.~B.~F.~M. 2010, \aap, 509,
  L6

\bibitem[{{Mittal} {et~al.}(2015){Mittal}, {Chen}, {Jang-Condell}, {Manoj},
  {Sargent}, {Watson}, \& {Lisse}}]{Mittal2015}
{Mittal}, T., {Chen}, C.~H., {Jang-Condell}, H., {et~al.} 2015, \apj, 798, 87

\bibitem[{{Mo{\'o}r} {et~al.}(2009){Mo{\'o}r}, {Apai}, {Pascucci},
  {{\'A}brah{\'a}m}, {Grady}, {Henning}, {Juh{\'a}sz}, {Kiss}, \&
  {K{\'o}sp{\'a}l}}]{Moor2009}
{Mo{\'o}r}, A., {Apai}, D., {Pascucci}, I., {et~al.} 2009, \apjl, 700, L25

\bibitem[{{Olofsson} {et~al.}(2009){Olofsson}, {Augereau}, {van Dishoeck},
  {Mer{\'{\i}}n}, {Lahuis}, {Kessler-Silacci}, {Dullemond}, {Oliveira},
  {Blake}, {Boogert}, {Brown}, {Evans}, {Geers}, {Knez}, {Monin}, \&
  {Pontoppidan}}]{Olofsson2009}
{Olofsson}, J., {Augereau}, J.-C., {van Dishoeck}, E.~F., {et~al.} 2009, \aap,
  507, 327

\bibitem[{{Olofsson} {et~al.}(2012){Olofsson}, {Juh{\'a}sz}, {Henning},
  {Mutschke}, {Tamanai}, {Mo{\'o}r}, \& {{\'A}brah{\'a}m}}]{Olofsson2012}
{Olofsson}, J., {Juh{\'a}sz}, A., {Henning}, T., {et~al.} 2012, \aap, 542, A90

\bibitem[{{Olofsson} {et~al.}(2016){Olofsson}, {Samland}, {Avenhaus},
  {Caceres}, {Henning}, {Mo{\'o}r}, {Milli}, {Canovas}, {Quanz}, {Schreiber},
  {Augereau}, {Bayo}, {Bazzon}, {Beuzit}, {Boccaletti}, {Buenzli}, {Casassus},
  {Chauvin}, {Dominik}, {Desidera}, {Feldt}, {Gratton}, {Janson}, {Lagrange},
  {Langlois}, {Lannier}, {Maire}, {Mesa}, {Pinte}, {Rouan}, {Salter},
  {Thalmann}, \& {Vigan}}]{Olofsson2016}
{Olofsson}, J., {Samland}, M., {Avenhaus}, H., {et~al.} 2016, \aap, 591, A108

\bibitem[{{Oudmaijer} {et~al.}(1992){Oudmaijer}, {van der Veen}, {Waters},
  {Trams}, {Waelkens}, \& {Engelsman}}]{Oudmaijer1992}
{Oudmaijer}, R.~D., {van der Veen}, W.~E.~C.~J., {Waters}, L.~B.~F.~M.,
  {et~al.} 1992, \aaps, 96, 625

\bibitem[{{Pavlov} {et~al.}(2008){Pavlov}, {Feldt}, \& {Henning}}]{Pavlov2008}
{Pavlov}, A., {Feldt}, M., \& {Henning}, T. 2008, in Astronomical Society of
  the Pacific Conference Series, Vol. 394, Astronomical Data Analysis Software
  and Systems XVII, ed. R.~W. {Argyle}, P.~S. {Bunclark}, \& J.~R. {Lewis}, 581

\bibitem[{{Pecaut} {et~al.}(2012){Pecaut}, {Mamajek}, \& {Bubar}}]{Pecaut2012}
{Pecaut}, M.~J., {Mamajek}, E.~E., \& {Bubar}, E.~J. 2012, \apj, 746, 154

\bibitem[{{Perrin} {et~al.}(2015){Perrin}, {Duchene}, {Millar-Blanchaer},
  {Fitzgerald}, {Graham}, {Wiktorowicz}, {Kalas}, {Macintosh}, {Bauman},
  {Cardwell}, {Chilcote}, {De Rosa}, {Dillon}, {Doyon}, {Dunn}, {Erikson},
  {Gavel}, {Goodsell}, {Hartung}, {Hibon}, {Ingraham}, {Kerley}, {Konapacky},
  {Larkin}, {Maire}, {Marchis}, {Marois}, {Mittal}, {Morzinski}, {Oppenheimer},
  {Palmer}, {Patience}, {Poyneer}, {Pueyo}, {Rantakyr{\"o}}, {Sadakuni},
  {Saddlemyer}, {Savransky}, {Soummer}, {Sivaramakrishnan}, {Song}, {Thomas},
  {Wallace}, {Wang}, \& {Wolff}}]{Perrin2015}
{Perrin}, M.~D., {Duchene}, G., {Millar-Blanchaer}, M., {et~al.} 2015, \apj,
  799, 182

\bibitem[{{Plavchan} {et~al.}(2009){Plavchan}, {Werner}, {Chen}, {Stapelfeldt},
  {Su}, {Stauffer}, \& {Song}}]{Plavchan2009}
{Plavchan}, P., {Werner}, M.~W., {Chen}, C.~H., {et~al.} 2009, \apj, 698, 1068

\bibitem[{{Press} {et~al.}(2007){Press}, {Teukolsky}, {Vetterling}, \&
  {Flannery}}]{Press2007}
{Press}, W.~H., {Teukolsky}, S.~A., {Vetterling}, W.~T., \& {Flannery}, B.~P.
  2007, Numerical Recipes 3rd Edition: The Art of Scientific Computing
  (Cambridge)

\bibitem[{{Roelfsema} {et~al.}(2010){Roelfsema}, {Schmid}, {Pragt}, {Gisler},
  {Waters}, {Bazzon}, {Baruffolo}, {Beuzit}, {Boccaletti}, {Charton}, {Cumani},
  {Dohlen}, {Downing}, {Elswijk}, {Feldt}, {Groothuis}, {de Haan}, {Hanenburg},
  {Hubin}, {Joos}, {Kasper}, {Keller}, {Kragt}, {Lizon}, {Mouillet}, {Pavlov},
  {Rigal}, {Rochat}, {Salasnich}, {Steiner}, {Thalmann}, {Venema}, \&
  {Wildi}}]{Roelfsema2010}
{Roelfsema}, R., {Schmid}, H.~M., {Pragt}, J., {et~al.} 2010, in \procspie,
  Vol. 7735, Ground-based and Airborne Instrumentation for Astronomy III,
  77354B

\bibitem[{{Schmid} {et~al.}(2017){Schmid}, {Bazzon}, {Milli}, {Roelfsema},
  {Engler}, {Mouillet}, {Lagadec}, {Sissa}, {Sauvage}, {Ginski}, {Baruffolo},
  {Beuzit}, {Boccaletti}, {Bohn}, {Claudi}, {Costille}, {Desidera}, {Dohlen},
  {Dominik}, {Feldt}, {Fusco}, {Gisler}, {Girard}, {Gratton}, {Henning},
  {Hubin}, {Joos}, {Kasper}, {Langlois}, {Pavlov}, {Pragt}, {Puget}, {Quanz},
  {Salasnich}, {Siebenmorgen}, {Stute}, {Suarez}, {Szul{\'a}gyi}, {Thalmann},
  {Turatto}, {Udry}, {Vigan}, \& {Wildi}}]{Schmid2017}
{Schmid}, H.~M., {Bazzon}, A., {Milli}, J., {et~al.} 2017, \aap, 602, A53

\bibitem[{{Schmid} {et~al.}(2012){Schmid}, {Downing}, {Roelfsema}, {Bazzon},
  {Gisler}, {Pragt}, {Cumani}, {Salasnich}, {Pavlov}, {Baruffolo}, {Beuzit},
  {Costille}, {Deiries}, {Dohlen}, {Dominik}, {Elswijk}, {Feldt}, {Kasper},
  {Mouillet}, {Thalmann}, \& {Wildi}}]{Schmid2012}
{Schmid}, H.-M., {Downing}, M., {Roelfsema}, R., {et~al.} 2012, in Society of
  Photo-Optical Instrumentation Engineers (SPIE) Conference Series, Vol. 8446,
  Society of Photo-Optical Instrumentation Engineers (SPIE) Conference Series,
  8

\bibitem[{{Schmid} {et~al.}(2006){Schmid}, {Joos}, \& {Tschan}}]{Schmid2006}
{Schmid}, H.~M., {Joos}, F., \& {Tschan}, D. 2006, \aap, 452, 657

\bibitem[{{Schneider} {et~al.}(2014){Schneider}, {Grady}, {Hines}, {Stark},
  {Debes}, {Carson}, {Kuchner}, {Perrin}, {Weinberger}, {Wisniewski},
  {Silverstone}, {Jang-Condell}, {Henning}, {Woodgate}, {Serabyn},
  {Moro-Martin}, {Tamura}, {Hinz}, \& {Rodigas}}]{Schneider2014}
{Schneider}, G., {Grady}, C.~A., {Hines}, D.~C., {et~al.} 2014, \aj, 148, 59

\bibitem[{{Schneider} {et~al.}(2016){Schneider}, {Grady}, {Stark}, {Gaspar},
  {Carson}, {Debes}, {Henning}, {Hines}, {Jang-Condell}, {Kuchner}, {Perrin},
  {Rodigas}, {Tamura}, \& {Wisniewski}}]{Schneider2016}
{Schneider}, G., {Grady}, C.~A., {Stark}, C.~C., {et~al.} 2016, \aj, 152, 64

\bibitem[{{Schneider} {et~al.}(1999){Schneider}, {Smith}, {Becklin}, {Koerner},
  {Meier}, {Hines}, {Lowrance}, {Terrile}, {Thompson}, \&
  {Rieke}}]{Schneider1999}
{Schneider}, G., {Smith}, B.~A., {Becklin}, E.~E., {et~al.} 1999, \apjl, 513,
  L127

\bibitem[{{Sch{\"u}ppler} {et~al.}(2015){Sch{\"u}ppler}, {L{\"o}hne}, {Krivov},
  {Ertel}, {Marshall}, {Wolf}, {Wyatt}, {Augereau}, \&
  {Metchev}}]{Schueppler2015}
{Sch{\"u}ppler}, C., {L{\"o}hne}, T., {Krivov}, A.~V., {et~al.} 2015, \aap,
  581, A97

\bibitem[{{Smith} \& {Terrile}(1984)}]{Smith1984}
{Smith}, B.~A. \& {Terrile}, R.~J. 1984, Science, 226, 1421

\bibitem[{{Stapelfeldt} {et~al.}(2004){Stapelfeldt}, {Holmes}, {Chen}, {Rieke},
  {Su}, {Hines}, {Werner}, {Beichman}, {Jura}, {Padgett}, {Stansberry},
  {Bendo}, {Cadien}, {Marengo}, {Thompson}, {Velusamy}, {Backus}, {Blaylock},
  {Egami}, {Engelbracht}, {Frayer}, {Gordon}, {Keene}, {Latter}, {Megeath},
  {Misselt}, {Morrison}, {Muzerolle}, {Noriega-Crespo}, {Van Cleve}, \&
  {Young}}]{Stapelfeldt2004}
{Stapelfeldt}, K.~R., {Holmes}, E.~K., {Chen}, C., {et~al.} 2004, \apjs, 154,
  458

\bibitem[{{Stolker} {et~al.}(2016){Stolker}, {Dominik}, {Avenhaus}, {Min}, {de
  Boer}, {Ginski}, {Schmid}, {Juhasz}, {Bazzon}, {Waters}, {Garufi},
  {Augereau}, {Benisty}, {Boccaletti}, {Henning}, {Langlois}, {Maire},
  {M{\'e}nard}, {Meyer}, {Pinte}, {Quanz}, {Thalmann}, {Beuzit}, {Carbillet},
  {Costille}, {Dohlen}, {Feldt}, {Gisler}, {Mouillet}, {Pavlov}, {Perret},
  {Petit}, {Pragt}, {Rochat}, {Roelfsema}, {Salasnich}, {Soenke}, \&
  {Wildi}}]{Stolker2016}
{Stolker}, T., {Dominik}, C., {Avenhaus}, H., {et~al.} 2016, \aap, 595, A113

\bibitem[{{Su} {et~al.}(2005){Su}, {Rieke}, {Misselt}, {Stansberry},
  {Moro-Martin}, {Stapelfeldt}, {Werner}, {Trilling}, {Bendo}, {Gordon},
  {Hines}, {Wyatt}, {Holland}, {Marengo}, {Megeath}, \& {Fazio}}]{Su2005}
{Su}, K.~Y.~L., {Rieke}, G.~H., {Misselt}, K.~A., {et~al.} 2005, \apj, 628, 487

\bibitem[{{Tamura} {et~al.}(2006){Tamura}, {Fukagawa}, {Kimura}, {Yamamoto},
  {Suto}, \& {Abe}}]{Tamura2006}
{Tamura}, M., {Fukagawa}, M., {Kimura}, H., {et~al.} 2006, \apj, 641, 1172

\bibitem[{{Thalmann} {et~al.}(2010){Thalmann}, {Grady}, {Goto}, {Wisniewski},
  {Janson}, {Henning}, {Fukagawa}, {Honda}, {Mulders}, {Min},
  {Moro-Mart{\'{\i}}n}, {McElwain}, {Hodapp}, {Carson}, {Abe}, {Brandner},
  {Egner}, {Feldt}, {Fukue}, {Golota}, {Guyon}, {Hashimoto}, {Hayano},
  {Hayashi}, {Hayashi}, {Ishii}, {Kandori}, {Knapp}, {Kudo}, {Kusakabe},
  {Kuzuhara}, {Matsuo}, {Miyama}, {Morino}, {Nishimura}, {Pyo}, {Serabyn},
  {Shibai}, {Suto}, {Suzuki}, {Takami}, {Takato}, {Terada}, {Tomono}, {Turner},
  {Watanabe}, {Yamada}, {Takami}, {Usuda}, \& {Tamura}}]{Thalmann2010}
{Thalmann}, C., {Grady}, C.~A., {Goto}, M., {et~al.} 2010, \apjl, 718, L87

\bibitem[{{Thalmann} {et~al.}(2013){Thalmann}, {Janson}, {Buenzli}, {Brandt},
  {Wisniewski}, {Dominik}, {Carson}, {McElwain}, {Currie}, {Knapp},
  {Moro-Mart{\'{\i}}n}, {Usuda}, {Abe}, {Brandner}, {Egner}, {Feldt}, {Golota},
  {Goto}, {Guyon}, {Hashimoto}, {Hayano}, {Hayashi}, {Hayashi}, {Henning},
  {Hodapp}, {Ishii}, {Iye}, {Kandori}, {Kudo}, {Kusakabe}, {Kuzuhara}, {Kwon},
  {Matsuo}, {Mayama}, {Miyama}, {Morino}, {Nishimura}, {Pyo}, {Serabyn},
  {Suto}, {Suzuki}, {Takami}, {Takato}, {Terada}, {Tomono}, {Turner},
  {Watanabe}, {Yamada}, {Takami}, \& {Tamura}}]{Thalmann2013}
{Thalmann}, C., {Janson}, M., {Buenzli}, E., {et~al.} 2013, \apjl, 763, L29

\bibitem[{{Thalmann} {et~al.}(2011){Thalmann}, {Janson}, {Buenzli}, {Brandt},
  {Wisniewski}, {Moro-Mart{\'{\i}}n}, {Usuda}, {Schneider}, {Carson},
  {McElwain}, {Grady}, {Goto}, {Abe}, {Brandner}, {Dominik}, {Egner}, {Feldt},
  {Fukue}, {Golota}, {Guyon}, {Hashimoto}, {Hayano}, {Hayashi}, {Hayashi},
  {Henning}, {Hodapp}, {Ishii}, {Iye}, {Kandori}, {Knapp}, {Kudo}, {Kusakabe},
  {Kuzuhara}, {Matsuo}, {Miyama}, {Morino}, {Nishimura}, {Pyo}, {Serabyn},
  {Suto}, {Suzuki}, {Takahashi}, {Takami}, {Takato}, {Terada}, {Tomono},
  {Turner}, {Watanabe}, {Yamada}, {Takami}, \& {Tamura}}]{Thalmann2011}
{Thalmann}, C., {Janson}, M., {Buenzli}, E., {et~al.} 2011, \apjl, 743, L6

\bibitem[{{Thalmann} {et~al.}(2008){Thalmann}, {Schmid}, {Boccaletti},
  {Mouillet}, {Dohlen}, {Roelfsema}, {Carbillet}, {Gisler}, {Beuzit}, {Feldt},
  {Gratton}, {Joos}, {Keller}, {Kragt}, {Pragt}, {Puget}, {Rigal}, {Snik},
  {Waters}, \& {Wildi}}]{Thalmann2008}
{Thalmann}, C., {Schmid}, H.~M., {Boccaletti}, A., {et~al.} 2008, in Society of
  Photo-Optical Instrumentation Engineers (SPIE) Conference Series, Vol. 7014,
  Society of Photo-Optical Instrumentation Engineers (SPIE) Conference Series,
  3

\bibitem[{{Trujillo} {et~al.}(2001){Trujillo}, {Aguerri}, {Cepa}, \&
  {Guti{\'e}rrez}}]{Trujillo2001}
{Trujillo}, I., {Aguerri}, J.~A.~L., {Cepa}, J., \& {Guti{\'e}rrez}, C.~M.
  2001, \mnras, 328, 977

\bibitem[{{van Leeuwen}(2007)}]{vanLeeuwen2007}
{van Leeuwen}, F. 2007, \aap, 474, 653

\bibitem[{{Wahhaj} {et~al.}(2007){Wahhaj}, {Koerner}, \&
  {Sargent}}]{Wahhaj2007}
{Wahhaj}, Z., {Koerner}, D.~W., \& {Sargent}, A.~I. 2007, \apj, 661, 368

\bibitem[{{Wyatt}(2008)}]{Wyatt2008}
{Wyatt}, M.~C. 2008, \araa, 46, 339

\end{thebibliography}

\end{document}